\documentclass[notitlepage,a4,10.5pt]{article}

\usepackage{graphicx}

\usepackage{amsmath}%
\usepackage{amsfonts}%
\usepackage{amssymb}%
\usepackage{mathrsfs}
\usepackage{amsthm}
\usepackage{float}

\usepackage{authblk}

\usepackage[left=2.5cm,right=2.5cm,top=3cm,bottom=3cm]{geometry} 

\usepackage{xcolor}

\newcommand{\mc}{\mathcal}
\newcommand{\cp}{\times}

\newcommand{\bol}{\boldsymbol}

\newcommand{\abs}[1]{\left\lvert{#1}\right\rvert}

\newcommand{\lr}[1]{\left({#1}\right)}

\newcommand{\p}{\partial}

\newcommand{\ti}[1]{\textit{#1}}
\newcommand{\tb}[1]{\textbf{#1}}

\begin{document}

\title{Maximum Entropy States of Collisionless Positron-Electron Plasma\\ in a Dipole Magnetic Field}
\author[1]{Naoki Sato} 
\affil[1]{Graduate School of Frontier Sciences, \protect\\ The University of Tokyo, Kashiwa, Chiba 277-8561, Japan \protect\\ Email: sato\_naoki@edu.k.u-tokyo.ac.jp}
\date{\today}
\setcounter{Maxaffil}{0}
\renewcommand\Affilfont{\itshape\small}

    \maketitle
    \begin{abstract}
    We are developing a positron-electron plasma trap based on a dipole magnetic field generated by a levitated superconducting magnet 
    to investigate the physics of magnetized plasmas with mass symmetry as well as antimatter components. 
    Such laboratory magnetosphere is deemed essential for the understanding of pair plasmas in astrophysical environments, 
    such as magnetars and blackholes, and represents a novel technology with potential applications in antimatter confinement 
    and development of coherent gamma-ray lasers. 
    The design of the device requires a preemptive analysis of the achievable self-organized steady states. 
    In this study, we construct a theoretical model describing 
    maximum entropy states 
    of a collisionless positron-electron plasma confined by a dipole magnetic field,   
    and demonstrate efficient confinement 
    of both species under a wide range of physical parameters by 
    analysing the effect of the three adiabatic invariants on the phase space distribution function.
    The theory is verified by numerical evaluation of spatial density, electrostatic potential, and toroidal rotation velocity 
    for each species in correspondence of the maximum entropy state. 
    \end{abstract}

\section{Introduction}

Pair plasmas consist of two species of charged particles having the same mass, but opposite electric charge. 
Such mass symmetry is in stark contrast with usual plasmas, where ions exhibit a significantly larger 
mass than the electron component, and it is the reason why pair plasmas are expected to possess
peculiar turbulence, stability, and fluctuation properties \cite{Stone2020}.  
Positron-electron plasmas are one example of naturally occurring pair plasmas, and can be found in astrophysical jets and magnetospheres 
of quasars \cite{Wardle,Hirotani,Warwick}, pulsars \cite{Sturrock}, and magnetars \cite{Beloborodov,Camilo,Duncan}, 
where positrons and electrons are formed by pair production.
More exotic pair plasmas, such as proton-antiproton plasmas, 
could occur as well in suitable astrophysical environments. 

Positron-electron plasma experiments based on magnetic mirrors have been considered in the past \cite{Tsy} and efforts are underway for their realization \cite{Higaki},
while pair plasmas involving oppositely charged fullerene ions have been produced in the laboratory \cite{Oohara,Kono}.   
Furthermore, positron-electron pair production is expected to occur during the disruptive phase of tokamaks \cite{Helander2003}.
At present, positron-electron plasma confinement schemes based on dipole magnetic fields represent an active area of research 
due to recent technological advancement in trap design \cite{Yos06,Saitoh2004,Boxer,Stanja18}, positron accumulation  \cite{Higaki2,Chen}, and positron injection \cite{Saitoh2015,Sten18}.
Furthermore, the inhomogeneity of a dipole magnetic field makes these magnetic configurations suitable to confine both neutral and nonnetutral plasmas, 
in contrast with magnetic traps relying on a straight homogeneous magnetic field, 
which are appropriate for nonneutral plasmas \cite{Malmberg,Dubin,Fajans}.

If realized, a positron-electron plasma based on a dipole magnetic field has several potential applications 
ranging from the experimental study of astrophysical systems such as astrophysical jets, quasars, and pulsars,  
to matter and antimatter confinement, potentially providing confinement to a large amount 
of charged particles (mainly depending on the properties of the matter/antimatter source) 
for a long time (300 seconds and beyond \cite{YosPRL}) without requiring external electric fields, 
and to development of coherent gamma-ray sources (gamma-ray lasers) that exploit 
the annihilation of positron-electron pairs. 

The target parameters for a  
laboratory positron-electron plasma are expected to ensure a high-degree of stability not seen 
in standard ion-electron plasmas \cite{Helander2014,Helander2016}.  
In such confinement regime,   
the gyroradius $r_{c}=mv_{\perp}/\abs{q}B$ is smaller than the Debye length    
$\lambda_D=\lr{\epsilon_0 k_BT/2ne^2}^{1/2}$, which is smaller than the system size $R$, i.e. $r_c<<\lambda_{D}<<R$. 
Here, $m$ denotes the particle mass, $v_{\perp}$ the velocity perpendicular to the magnetic field $\bol{B}$, $q$ the electric charge, $B$ the modulus of the magnetic field, $\epsilon_0$ the vacuum permittivity, $k_B$ the Boltzmann constant, $T$ the temperature, $n$ the spatial density, and $e=\abs{q}$. 
For a positron-electron plasma with $B\sim 1\, T$, $n\sim 10^{12}\, m^{-3}$, 
$k_B T\sim 10\, eV$, and $R\sim 1 \,m$, one obtains $r_c\sim 10^{-5}\,m$ and $\lambda_D\sim 2\,10^{-2}\,m$. 
Furthermore, the plasma parameter $\Lambda=4\pi n\lambda_D^3=(1/4\sqrt{2\pi})\lr{r_d/r_C}^{3/2}\sim 10^8>>1$ is large, implying that the typical distance among  particles $r_{d}=n^{-1/3}\sim 10^{-4}\,m$ is much larger than the distance $r_{C}=e^2/4\pi\epsilon_0 k_BT\sim 10^{-10}\,m$ at which the Coulomb energy becomes  comparable with the average kinetic energy. This makes the system weakly coupled, and diffusive (entropy maximizing) processes are dominated by collective electromagnetic fluctuations rather than localized Coulomb collisions. 

At present, we are developing a positron-electron plasma trap 
based on a dipole magnetic field generated by a levitated superconducting coil 
operating in the plasma regime described above. 
This requires a preemptive analysis of
the achievable plasma confinement. 
The aim of this study is thus to explore the nature of 
maximum entropy states 
that are self-organized by the plasma, and to assess the degree of confinement of both species. 
In this context, a maximum entropy state is defined as the distribution function of largest entropy that is compatible with the macroscopic constraints affecting the system, such as conservation of total particle number, total energy, or total magnetic moment. 
Since a positron-electron system is formally 
analogous to any two-species plasma, such as  
an ion-electron plasma, thermal equilibria in a positron-electron plasma can be inferred from those of a two-species plasma. 
It is therefore important to stress that the core issue examined in the present paper is not the study of two-species thermal equilibria, which is an established matter, 
but the elucidation of the effect of the conservation of adiabatic invariants on self-organized maximum entropy states in two-species plasmas, a problem that is not discussed in the literature.

Once built, the positron-electron plasma trap will be used to study      
both waves and transport phenomena, which are expected to  be in the $1\,kHz$ and $1\,Hz$ frequency range  respectively based on previous experimental data from the RT-1 device \cite{YosPRL}. Hence, 
we aim at achieving confinement on time scales $\tau_c$ of the order of $1\,s$.
Since in the present setting the Coulomb scattering frequency is of the order $\nu_C\sim 1\,Hz$, 
the plasma is expected to be quite collisionless, while 
the main mechanism increasing the entropy of the system is given by collective electromagnetic fluctuations. 
Hence, we shall develop the theory by assuming that Coulomb collisions 
can be neglected. 

It should be emphasized that accumulating $10^{11}$-$10^{12}$ positrons in a volume $\Omega\sim 1\,m^{3}$ requires the development of positron accumulation and injection technologies. 
In the present trap design, 
positrons will be produced by a $10\,eV$ pulsed positron source 
located at 
The National Institute of Advanced Industrial Science \cite{Higaki2},   and progressively accumulated into a buffer-gas trap until a total positron number $N_p\sim 10^{11}$ is reached. The positrons will then be released into the dipole trap over a time interval of $10\,\mu s$, reaching a particle density in the target range $10^{11}$-$10^{12}\,m^{-3}$. 
Efficient injection schemes are also being developed to minimize losses at the injection phase.

In addition to the technological hurdles listed above, other factors, such as particle loss associated with turbulent transport and plasma instability, may affect the quality of confinement. 
These aspects, which have been investigated in dipole geometry 
within the frameworks of MHD and gyrokinetics  (see e.g. \cite{Garnier,Kesner,Kobayashi}),
will not be addressed in the present paper, since its scope is limited at  establishing the existence of stable maximum entropy equilibria toward which the positron-electron  plasma is expected to converge under ideal conditions. 

In an inhomogeneous magnetic field such as a dipole magnetic field, 
the properties of the equilibrium states are strongly dependent on the
presence of adiabatic invariants \cite{LandauMec}. 
When the time scale of collective electromagnetic fluctuations 
within the plasma are longer than the period of 
cyclotron gyration, bounce motion, or toroidal drift, 
the corresponding adiabatic invariant (magnetic moment $\mu$, bounce action $J_{\parallel}$, or magnetic flux $\Psi$) is conserved 
by the dynamics of a single charged particle. 
Indeed, the essential feature of adiabatic invariants is that they remain approximately constant as long as   perturbations acting on a dynamical system are slow compared with the period of adiabatic motion.
This results in a set of constraints on the equilibrium distribution function of the system,
which departs from a standard Maxwell-Boltzmann distribution \cite{Hasegawa1990,YosPPCF,YosPTEP}.
To understand how this occurs, it is useful to consider the limiting case in which
the value of one adiabatic invariant, say $\mu$, is preserved exactly. 
Then, 
on each phase space submanifold corresponding to a level set of $\mu$ (a $\mu$-leaf) 
the effective phase space measure
is reduced to $Bdxdydz dv_{\parallel}$, where $\lr{x,y,z}$ are Cartesian coordinates
and $v_{\parallel}$ denotes the velocity along the magnetic field. 
If $f_{\mu}\lr{x,y,z,v_{\parallel}}$ is the distribution function on 
a $\mu$-leaf, entropy maximizing processes such as cross-field diffusion 
lead to a flattening of $f_{\mu}$, with a corresponding 
relaxed spatial density
$n=B\int f_{\mu}dv_{\parallel}d\mu\sim B$, which is inhomogeneous for an inhomogeneous $B$ (for further details see \cite{YosPTEP,Sato2015,Sato2016}). 
Therefore, a 
statistical description of 
maximum entropy states 
must take into account 
the nontrivial role played by adiabatic invariants in shaping the distribution function of each particle species.

The present paper is organized as follows. 
In section 2, we derive the maximum entropy distribution functions of positrons and electrons 
by taking into account the conservation of the first adiabatic invariant,  
and obtain the corresponding form of the Poisson equation for the electrostatic potential.
In sections 3 and 4, we study the effect of the second and third adiabatic invariants 
on positron-electron plasma maximum entropy states. 
In section 5, we report spatial densities and electrostatic potential obtained by numerical solution of the Poisson equation for the electrostatic potential, and demonstrate efficient confinement of both species. 
In section 6, we study the toroidal rotation velocity profile of the positron-electron plasma, and find that it departs from the rigid rotation occurring in 
magnetic traps relying on a straight homogeneous magnetic field. 
Concluding remarks are given in section 7. 

We remark that the theory developed in the present paper holds provided that the macroscopic constraints defining each statistical ensemble (total particle number, total energy, total magnetic moment, and so on) are preserved throughout the 
relaxation 
of the system. In particular, 
the
period of the slowest adiabatic motion 
must be considerably smaller than the
typical time scale of the fastest 
turbulent fluctuations in the system. 
However, in real experiments particles and energy will be lost via different mechanisms, such as by interaction with the vessel boundary, and the degree of conservation of  adiabatic invariants 
will not be perfect. 
Therefore, the theory will be quantitatively accurate only if such losses are not too large. Nevertheless, we expect the theory to remain qualitatively consistent even in the presence of large losses since the surviving (trapped component) of the plasma should eventually converge toward the derived 
maximum entropy states.

Finally, notice that 
 the present theory does not make any predictions
on the nature of electromagnetic fluctuations. It only provides information on the 
maximum entropy state 
that results from the action of these fluctuations, which act to increase the entropy of the system.
Hence, although the model cannot describe transient phenomena such as waves, transport, or diffusion,
it is sufficient to describe the ‘final’ state of the system, i.e. the equilibrium toward which the plasma
tends to settle over a sufficiently long time interval.













\section{Spatial density and electrostatic potential in magnetized positron-electron plasma}
In the following lower indexes $p$ and $e$ will be used to specify positrons and electrons respectively.
In a static equilibrium, the electrostatic  potential $\Phi$ is determined by the Poisson equation
\begin{equation}
\Delta\Phi=-\frac{\rho}{\epsilon_0}.
\label{Mx1}
\end{equation}
Here, $\rho$ is the electric charge density and $\epsilon_0$ the vacuum permittivity.
In a positron-electron plasma, the charge density can be expressed as
\begin{equation}
\rho=e\lr{n_p-n_e},
\end{equation}
where $e$ is the positron charge, $-e$ the electron charge,
$n_p$ the positron number density, and $n_e$ the electron number density.   
Denoting with $f_p$ the probability distribution function of positrons,
with $f_e$ the probability distribution function of electrons, 
with $N_p$ the total number of positrons
, and with $N_e$ the total number of electrons, the number densities $n_p$ and $n_e$ can be evaluated as
\begin{equation}
n_p=N_p\int_{\mathbb{R}^3}f_p\,d^3p,~~~~n_e=N_e\int_{\mathbb{R}^3}f_e\,d^3p.\label{npne}
\end{equation}
In this notation $d^3p$ denotes the volume element in momentum space. 

Let $\Omega\subset\mathbb{R}^3$ denote the spatial volume occupied by the positron-electron plasma,   
$\Pi=\mathbb{R}^3\cp\Omega$ the phase space of the system, and $d\Pi=d^3pd^3x$ the phase space measure. 
Notice that the phase space measure $d\Pi$ is invariant due to Liouville's theorem arising from the underlying Hamiltonian structure, 
and therefore serves as the natural measure with respect to which the ergodic hypothesis of statistical mechanics is enforced. Furthermore, it should be emphasized that the measure $d\Pi$ remains invariant for any reduced subsystem of charged particle dynamics, such as guiding center dynamics. 
Next, we consider a guiding center plasma such that the first adiabatic invariant $\mu$ is a constant of motion of (isolated) charged particle dynamics. 
Here, the magnetic moment $\mu$ \cite{Haz,Cary,North} is defined as 
\begin{equation}
\mu=\frac{m\abs{\bol{v}_{\perp}-\bol{v}_d}^2}{2B}=\frac{m{v}_c^2}{2B},\label{mu}
\end{equation}
where $\bol{v}_{\perp}$ is the particle velocity across the magnetic field, $\bol{v}_d$ the component of $\bol{v}$ independent of the cyclotron phase $\vartheta_c$ and comprising the guiding center drifts, $\bol{v}_c=\bol{v}_{\perp}-\bol{v}_d$ the cyclotron velocity with modulus $v_c$, $B$ the modulus of the magnetic field $\bol{B}$, and $m$ the particle mass (which is the same for both positrons and electrons). 

For the first adiabatic invariant to be preserved, the time scale $T_f$ of fluctuations in particle energy caused by 
electromagnetic turbulence must be longer than the time scale of cyclotron motion $T_c$, i.e. 
\begin{equation}
T_f>>T_c=\frac{2\pi}{\omega_c}=\frac{2\pi m}{eB},\label{tftc}
\end{equation}
where $\omega_c$ denotes the cyclotron frequency. In the following, we shall therefore assume that the condition \eqref{tftc} holds. 
We also recall that the conservation of the quantity \eqref{mu}, usually referred to as lowest order magnetic moment, rests on the additional requirement that the wavelength $\lambda$ of any fluctuation is longer than the Larmor radius $\rho_c$, i.e. $\rho_c<<\lambda$. 
In this setting, the guiding center Hamiltonian functions for positron and electrons have expressions
\begin{equation}
H_{p}=\frac{1}{2}mv_{\parallel}^2+\mu B+e\Phi,~~~~H_e=\frac{1}{2}mv_{\parallel}^2+\mu B-e\Phi.\label{Hgc}
\end{equation}
Here $v_{\parallel}$ is the velocity component along the magnetic field.
Observe that the drift velocity $\bol{v}_d$ does not contribute to the energies \eqref{Hgc} under the assumption that the corresponding kinetic energy is small compared to the other terms. 
This is true, for example, when the  potential energy $e\Phi/k_BT\sim\epsilon$, with $\epsilon=\rho_c/L$, $L$ the characteristic scale lenght of the magnetic field, and $k_BT$ a characteristic plasma temperature,  represents a small perturbation of the particle energy (for additional details on this point see discussion at the end of section 4 or \cite{Cary}). 
Then, the Shannon entropies $S_p$ and $S_e$, the total probabilities $P_p$ and $P_e$, the total energies $E_p$ and $E_e$, and the total magnetic moments $M_p$ and $M_e$ of the system can be written as
\begin{subequations}
\begin{align}
S_p=&-\int_{\Pi}N_pf_p\log \lr{N_pf_p}\,d\Pi,~~~~S_e=-\int_{\Pi}N_ef_e\log \lr{N_ef_e}\,d\Pi,\label{Sh}\\
P_p=&\int_{\Pi}f_{p}d\Pi,~~~~P_e=\int_{\Pi}f_{e}d\Pi,\\
E_p=&N_p\int_{\Pi}f_pH_p\,d\Pi,~~~~E_e=N_e\int_{\Pi}f_eH_e\,d\Pi\\
M_p=&N_p\int_{\Pi}f_p\mu\,d\Pi,~~~~M_e=N_e\int_{\Pi}f_{e}\mu\,d\Pi.
\end{align}
\end{subequations}
Under the ansatz \eqref{Sh} for the entropy measure associated with each particle species,  
the thermodynamic equilibrium of the system can be obtained by maximization of entropy under the 
constraint imposed by the constancy of $P_p$, $P_e$, $E_p$, $E_e$, $M_p$, and $M_e$ (for further details on this approach see \cite{YosPPCF,YosPTEP}). 
Introducing Lagrange multipliers $\alpha_p$, $\alpha_e$, $\beta_p$, $\beta_e$, $\gamma_p$, and $\gamma_e$,   
we therefore define the target functional 
\begin{equation}
F\left[f_p,f_e\right]=S_p+S_e-\alpha_p N_p P_p-\alpha_e N_e P_e-\beta_pE_p-\beta_eE_e-\gamma_p M_p-\gamma_e M_e.\label{TF}
\end{equation}
The Euler-Lagrange equations obtained by variation of \eqref{TF} with respect to $f_p$ and $f_e$ are:
\begin{subequations}
\begin{align}
f_p=&A_p\exp\left\{-\beta_pH_p-\gamma_p\mu\right\},\label{fpfea}\\
f_e=&A_e\exp\left\{-\beta_eH_e-\gamma_e\mu\right\},
\end{align}\label{fpfe}
\end{subequations}
where $A_p=N_p^{-1}\exp\left\{-1-\alpha_p\right\}$ and $A_e=N_e^{-1}\exp\left\{-1-\alpha_e\right\}$ are normalization factors such that $\int_{\Pi}f_p\,d\Pi=\int_{\Pi}f_e\,d\Pi=1$. 
In the same way $\beta_p$ and $\beta_e$ 
represent the 
characteristic inverse temperatures of the two particle species, 
the Lagrange multipliers $\gamma_p$ and $\gamma_e$ 
can be interpreted as chemical potentials 
describing the macroscopic energy changes $\beta_p dE_p=\gamma_p dM_p$ and $\beta_e dE_e=\gamma_e dM_e$ occurring when magnetic moments $dM_p,dM_e$ are added to the system.

Next, it is useful to explain why Coulomb collisions have been neglected in the derivation of  the collisionless equilibria \eqref{fpfe}.  
First, recall that in the present setting the (collective) electric potential $\Phi$ 
changes over a time scale $T_f$ that is much longer than the cyclotron time scale $T_c$. 
This ensures that the individual magnetic moments of the particles remain constant, unless local Coulomb collisions occur. 
For the plasma regime 
under consideration, the frequency $\nu_C$ of such collisions can be estimated \cite{Goldston} as $\nu_C\sim 5\,10^{-11}nT^{-3/2}\sim 1\,Hz$, where $n=10^{12}\,m^{-3}$ is the particle density and $T=10\,eV$ the temperature. 
Hence, over the confinement time scales $\tau_c\sim 1/\nu_C\sim1\,s$ considered in the model they are negligible, 
their effect being felt only over longer time intervals where particles undergo mostly small deflections with small changes in their kinetic energies. Indeed, recall that the distance $r_C\sim 10^{-10}\,m$ at which Coulomb interactions are dominant is smaller than the typical  particle distance $r_d\sim 10^{-4}\,m$. 
We also remark that if a net transfer of total energy or total magnetic moment occurs 
between positrons and electrons, 
only the sums $E_p+E_e$ and $M_p+M_e$ are preserved. Such scenario can be taken into account by setting $\beta=\beta_e=\beta_p$ and $\gamma=\gamma_p=\gamma_e$. 

The expressions \eqref{fpfe} for the distribution functions $f_p$ and $f_e$ can now be used to determine the spatial number densities $n_p$ and $n_e$ according to \eqref{npne}. To evaluate the integral in \eqref{npne}, 
the phase space measure $d\Pi$ must be expressed in a more convenient set of magnetic coordinates. 
To this end, we restrict our attention to magnetic fields of the type
\begin{equation}
\bol{B}=\nabla\Psi\cp\nabla\varphi,\label{B1}
\end{equation}
where $\Psi$ denotes the flux function and $\varphi$ the toroidal angle. Introducing a length coordinate $\ell$ 
along magnetic field lines with tangent vector $\p_{\ell}=\bol{B}/B$, the functions $\lr{\ell,\Psi,\varphi}$ 
define a set of curvilinear coordinates with Jacobian
\begin{equation}
\nabla\ell\cdot\nabla\Psi\cp\nabla\varphi=B\nabla\ell\cdot \p_\ell=B.\label{Jx}
\end{equation}
Next, decompose the  particle velocity as
\begin{equation}
\bol{v}=\bol{v}_{\parallel}+\bol{v}_c+\bol{v}_d,\label{parv}
\end{equation}
where $\bol{v}_{\parallel}=v_{\parallel}\bol{B}/B$ is the velocity component along $\bol{B}$, $\bol{v}_c$ the cyclotron velocity such that $mv_c^2=2\mu B$ (recall \eqref{mu}), and $\bol{v}_d=\bol{v}_d\lr{\bol{x}}$ the particle drift velocity across $\bol{B}$. 
Notice that 
\eqref{parv} represents the
velocity of a charged particle, and not the guiding center velocity. Furthermore, the drift velocity $\bol{v}_d$  comprising $\bol{E}\cp\bol{B}$, gradient, and curvature drifts is treated as a spatial function, which is expected to be a valid approximation in a time-independent setting since in a vacuum field gradient and curvature drifts can be expressed as $2k_BT\bol{B}\cp\bol{\kappa}/qB^2$, with $\bol{\kappa}$ the field curvature, $k_BT$ the temperature, and $q=\pm e$ the relevant electric charge. 
Let $r$ denote the cylindrical radius,  $\vartheta_c$ the phase of the cyclotron gyration, and $\bol{A}=\Psi\nabla \varphi$ the vector potential associated with the magnetic field \eqref{B1}. 
Then, 
the particle momentum $\bol{p}=m\bol{v}+q\bol{A}$ 
can be decomposed on the orthonormal set of basis vectors 
\begin{equation}
\lr{\p_{\ell},\frac{\nabla\Psi}{\abs{\nabla\Psi}},r\nabla\varphi},
\end{equation}
as 
\begin{equation}
\bol{p}=p_{\parallel}\p_{\ell}+p_{\Psi}\frac{\nabla\Psi}{\abs{\nabla\Psi}}+\frac{p_{\varphi}}{r}r\nabla\varphi,
\end{equation}
with 
\begin{subequations}
\begin{align}
p_{\parallel}=&mv_{\parallel},\\
p_{\Psi}=&m\lr{v_c\sin\vartheta_c+\frac{\bol{v}_d\cdot\nabla\Psi}{\abs{\nabla\Psi}}},\\
p_{\varphi}=&mr\lr{v_c\cos\vartheta_c+r\bol{v}_d\cdot\nabla\varphi}+q\Psi.
\end{align}\label{p}
\end{subequations}
It follows that at each point $\bol{x}\in\Omega$, 
\begin{equation}
d^3p=dp_{\parallel}dp_{\Psi}d\lr{\frac{p_{\varphi}}{r}}=m^2Bdv_{\parallel}d\vartheta_c d\mu.
\label{Jp} 
\end{equation}
Combining \eqref{Jx}, \eqref{p}, and \eqref{Jp}, we thus arrive at   
\begin{equation}
d\Pi=m^2d\ell dv_{\parallel}d\varphi d\Psi d\vartheta_c d\mu.\label{dpi}
\end{equation}
Recalling \eqref{npne} and assuming $\beta_p\geq0$, $\beta_e\geq0$, $\gamma_p\geq 0$, and $\gamma_e\geq 0$, the densities $n_p$ and $n_e$ therefore have expressions,
\begin{subequations}
\begin{align}
n_p&=m^2N_p\int_0^\infty d\mu\int_0^{2\pi}d\vartheta_c\int_{-\infty}^{\infty}dv_{\parallel}A_pB\exp\left\{
-\beta_p\lr{\frac{1}{2}mv_{\parallel}^2+\mu B+e\Phi}-\gamma_p\mu
\right\}=\sigma_p\frac{B\exp\left\{-\beta_pe\Phi\right\}}{\gamma_p+\beta_pB},\\
n_e&=m^2N_e\int_0^\infty d\mu\int_0^{2\pi}d\vartheta_c\int_{-\infty}^{\infty}dv_{\parallel}A_eB\exp\left\{
-\beta_e\lr{\frac{1}{2}mv_{\parallel}^2+\mu B-e\Phi}-\gamma_e\mu
\right\}=\sigma_e\frac{B\exp\left\{\beta_ee\Phi\right\}}{\gamma_e+\beta_eB},
\end{align}\label{npne2}
\end{subequations}
where we defined the constants
\begin{equation}
\sigma_p=\frac{\lr{2\pi m}^{\frac{3}{2}}A_pN_p}{\sqrt{\beta_p}},~~~~\sigma_e=\frac{\lr{2\pi m}^{\frac{3}{2}}A_eN_e}{\sqrt{\beta_e}}.
\end{equation}
Substituting these expression into equation \eqref{Mx1}, 
one thus obtains a second order nonlinear partial differential equation
\begin{equation}
\Delta\Phi=-\frac{e\lr{2\pi m}^{\frac{3}{2}}}{\epsilon_0}B\lr{\frac{A_pN_p}{\sqrt{\beta_p}}\frac{\exp\left\{-\beta_pe\Phi\right\}}{\gamma_p+\beta_p B}-\frac{A_eN_e}{\sqrt{\beta_e}}\frac{\exp\left\{\beta_ee\Phi\right\}}{\gamma_e+\beta_eB}}.\label{PhiEq}
\end{equation}
governing the spatial behavior of the electrostatic potential $\Phi$ 
within the magnetized positron-electron plasma under the effect of the magnetic field \eqref{B1}. 

It is useful to consider equation \eqref{PhiEq} in the following limits. First, if $N_e>>N_p$ the system approaches a pure electron plasma. In such case equation \eqref{PhiEq} reduces to
\begin{equation}
\Delta\Phi=\frac{e\lr{2\pi m}^{\frac{3}{2}}A_eN_e}{\epsilon_0\sqrt{\beta_e}}\frac{B\exp\left\{\beta_ee\Phi\right\}}{\gamma_e+\beta_eB}.\label{PhiEqPureEl}
\end{equation}
Secondly, if $N_p=N_e=N$, the inverse temperature of the positron plasma
equals that of the electron plasma
$\beta_p=\beta_e=\beta$, and the Lagrange multipliers associated with conservation of magnetic moment satisfy  $\gamma_p=\gamma_e=\gamma$ as well,  
equation \eqref{PhiEq} becomes
\begin{equation}
\Delta\Phi=-\frac{e\lr{2\pi m}^{\frac{3}{2}}N}{\epsilon_0\sqrt{\beta}}\frac{B}{\gamma+\beta B}\lr{A_p\exp\left\{-\beta e\Phi\right\}-A_e\exp\left\{\beta e\Phi\right\}}.\label{PhiEq2}
\end{equation}
Finally, if sufficiently long time scales are considered, the conservation of the total magnetic moments $M_p$ and $M_e$ is expected to break down. This scenario corresponds to the limit $\gamma_p=\gamma_e=0$, which gives
\begin{equation}
\Delta\Phi=-\frac{e\lr{2\pi m}^{\frac{3}{2}}}{\epsilon_0}\lr{\frac{A_pN_p}{\beta_p^{\frac{3}{2}}}\exp\left\{-\beta_pe\Phi\right\}-\frac{A_eN_e}{\beta_e^{\frac{3}{2}}}\exp\left\{\beta_ee\Phi\right\}}.\label{PhiEqM}
\end{equation}
By further demanding $N_e>>N_p$, one recovers the Liouville equation
\begin{equation}
\Delta\Phi=\frac{eA_eN_e}{\epsilon_0}\lr{\frac{2\pi m}{\beta_e}}^{\frac{3}{2}}\exp\left\{\beta_ee\Phi\right\}.\label{PhiEqL}
\end{equation}

We conclude this section by observing that the departure from Maxwell-Boltzmann statistics occurring in equilibria such as \eqref{fpfe} implies that the physical plasma temperatures (which differ from the temperatures $T_p$ and $T_e$ associated with the Lagrange multipliers $\beta_p=\lr{k_BT_p}^{-1}$ and $\beta_e=\lr{k_BT_e}^{-1}$) are not spatially uniform.  Indeed, integrals of the type
\begin{equation}
k_BT_{p\perp}\lr{\bol{x}}=\frac{\int_{\mathbb{R}^3}\mu Bf_p\,d^3p}{\int_{\mathbb{R}^3}f_p\,d^3p},~~~~k_BT_{p\parallel}\lr{\bol{x}}=\frac{\int_{\mathbb{R}^3}\frac{m}{2}v_{\parallel}^2f_p\,d^3p}{\int_{\mathbb{R}^3}f_p\,d^3p},
\end{equation}
will generally exhibit
a spatial dependence caused by the inhomogeneity of the magnetic field. 
Here, $T_{p\perp}$ and $T_{p\parallel}$ are the perpendicular and parallel temperatures of the positron plasma. 
For example, using equations \eqref{fpfea} and \eqref{Jp}, one obtains
\begin{equation}
k_BT_{p\perp}\lr{\bol{x}}=\frac{B\lr{\bol{x}}}{\beta_p B\lr{\bol{x}}+\gamma_p}.
\end{equation}

\section{The effect of bounce motion}
In this section, we consider a regime of plasma such that charged particles preserve 
both the first adiabatic invariant $\mu$ and 
the second adiabatic invariant (bounce action) $J_{\parallel}$, which is defined by 
\begin{equation}
J_{\parallel}=\frac{m}{2\pi}\int_{a}^{b}v_{\parallel}^{\ast}\,ds.\label{J1}
\end{equation}
Here, $a$ and $b$ denote the bouncing points along a field line with line element $ds$, while
\begin{equation}
v_{\parallel}^{\ast}\lr{E,\mu,\ell,\Psi,\varphi}=\sqrt{\frac{2}{m}\lr{E-\mu B-q\Phi}},
\end{equation}
represents the parallel velocity as a function of particle energy $E$, magnetic moment $\mu$,
and spatial position $\lr{\ell,\Psi,\varphi}$.
The conservation of the second adiabatic invariant \eqref{J1} 
requires that 
time scale $T_f$ of energy fluctuations caused by electromagnetic turbulence 
is longer than the 
time scale of bounce motion $T_b$, i.e. 
\begin{equation}
T_f>>T_b=\frac{2\pi}{\omega_b},\label{tftb}
\end{equation}
 where $\omega_b$ denotes the bounce frequency. 
The (half) period $T_b$ of the bounce oscillation can be written as
\begin{equation}
T_b
=\int_a^b\frac{ds}{v_{\parallel}^{\ast}}.\label{Tb}
\end{equation}
In the following we shall therefore assume that both \eqref{tftc} and \eqref{tftb} hold.

Next, observe that the bounce averaged  kinetic energy $\langle K_{\parallel}\rangle_b$ along the magnetic field
can be evaluated as
\begin{equation}
\langle K_{\parallel}\rangle_b=\frac{m}{2T_b}\int_{t_a}^{t_b}{v_{\parallel}^{\ast 2}}dt=\frac{m\omega_b}{4\pi}\int_a^bv_{\parallel}^{\ast}ds=\frac{1}{2}\omega_b J_{\parallel},\label{Kpar}
\end{equation}
where $t_a$ and $t_b$ are the instants at which the particle position reaches the bounce points $a$ and $b$, while  $\langle~\rangle_b$ denotes averaging over a bounce oscillation.

The actual magnetic field strength within the planned positron-electron trap will not be symmetric under vertical reflections $z\rightarrow -z$ due to 
a support coil placed at the top of the device to keep the
main superconducting coil generating the dipole magnetic field levitated. Therefore, 
an accurate model of the trap would need to take into account such asymmetry. 
This problem will not be considered here to simplify the analysis, 
and a pure dipole mangnetic field will be assumed. 
Due to the axial symmetry and reflection symmetry of the dipole magnetic field strength $B$, 
one has $a=-b$ as well as $\p v_{\parallel}^{\ast}/\p\varphi=0$.
Next, suppose that the number of positrons equals the number of electrons, $N_p=N_e$.  
Then, we may assume the electric potential energy $q\Phi$ to be small compared to the 
energy stored in the cyclotron gyration, $\mu B>>\abs{q\Phi}$. In this case, equations 
\eqref{J1} and \eqref{Tb} can be simplified to
\begin{equation}
J_{\parallel}=\frac{\sqrt{2m}}{\pi}\int_0^{b\lr{E,\mu,\Psi}}\sqrt{E-\mu B\lr{s,\Psi}}\,ds,~~~~T_b=\sqrt{2m}\int_0^{b\lr{E,\mu,\Psi}}\frac{ds}{\sqrt{E-\mu B\lr{s,\Psi}}},\label{J2Tb2}
\end{equation}
where the bouncing point $b\lr{E,\mu,\Psi}>0$ is given as the positive solution of the equation $E=\mu B\lr{b,\Psi}$, provided that such solution exists. 
Notice that both bounce action $J_{\parallel}$ and bounce frequency $\omega_b=2\pi/T_b$ are independent of  particle charge $q$ according to  \eqref{J2Tb2}, and therefore they  
have the same expression for both positrons and electrons. 
Now observe that, using the identity \eqref{Kpar} for the kinetic energy along the magnetic field, the positron energy and the electron energy can be approximated as
\begin{equation}
\tilde{H}_p=\frac{1}{2}\omega_b J_{\parallel}+\mu B+e\Phi,~~~~\tilde{H}_e=\frac{1}{2}\omega_b J_{\parallel}+\mu B-e\Phi.\label{HpHeJ}
\end{equation}
The errors $H_p-\tilde{H}_p$ and $H_e-\tilde{H}_e$ committed in replacing the energies $H_p, H_e$ with $\tilde{H}_p,\tilde{H}_e$ evidently goes to zero when averaged over a bounce oscillation, $\langle H_p-\tilde{H}_p\rangle_b=\langle H_e-\tilde{H}_e\rangle_b=0$.
In the following we shall be concerned with relaxation time scales $\tau$ longer than the bounce period, $\tau>>T_b$, 
and use the approximate expressions 
\eqref{HpHeJ} for the particles energies. As it will be shown later, this approach simplifies calculations involving the particles distribution functions.

The conservation of the second adiabatic invariant gives rise to a macroscopic constraint on the statistical behavior of the system, 
which can be represented by adding the total bounce actions $\mc{J}_p$ and $\mc{J}_e$ given by
\begin{equation}
\mc{J}_{p}=N_p\int_{\Pi}{f_pJ_{\parallel}\,d\Pi},~~~~\mc{J}_{e}=N_e\int_{\Pi}{f_eJ_{\parallel}\,d\Pi},
\end{equation}
in the target functional \eqref{TF} for the entropy principle.
Introducing Lagrange multipliers $\zeta_p$ and $\zeta_e$, the resulting expressions for 
the distribution functions $f_p$ and $f_e$ are
\begin{subequations}
\begin{align}
f_p=&A_p\exp\left\{-\beta_p \tilde{H}_p-\gamma_p\mu-\zeta_p J_{\parallel}\right\},\\
f_e=&A_e\exp\left\{-\beta_e \tilde{H}_e-\gamma_e\mu-\zeta_e J_{\parallel}\right\}.
\end{align}
\end{subequations}
Observe that $H_p$ and $H_e$ have been replaced with the approximated values $\tilde{H}_p$ and $\tilde{H}_e$. 
Our next goal is to evaluate the spatial particle densities \eqref{npne}, 
so that Poisson's equation \eqref{Mx1} can be applied to compute $\Phi$.
To this end, the phase space measure $d\Pi$ must be expressed in a new set of coordinates
that is appropriate to carry out integrals in momentum space. First, consider the 
change of variables $\lr{\ell,v_{\parallel},\varphi,\Psi,\theta_c,\mu}\rightarrow\lr{\ell,E,\varphi,\Psi,\theta_c,\mu}$ with $E=mv_{\parallel}^2/2+\mu B+q\Phi$. Recalling \eqref{dpi}, we thus have
\begin{equation}
d\Pi=m^2d\ell dv_{\parallel} d\varphi d\Psi d\vartheta_c d\mu=\frac{m}{v_{\parallel}^{\ast}}d\ell dE d\varphi d\Psi d\vartheta_c d\mu.
\end{equation}
Next, perform the transformation $\lr{\ell,E,\varphi,\Psi,\theta_c,\mu}\rightarrow\lr{\ell,J_{\parallel},\varphi,\Psi,\theta_c,\mu}$. In order to express the phase space measure in terms of the new coordinates, we must compute the partial derivative
\begin{equation}
\begin{split}
\frac{\p J_{\parallel}}{\p E}=&\lim_{dE\rightarrow 0}\frac{m}{\pi dE}\left[\int_0^{b\lr{E+dE,\mu,\Psi}}v_{\parallel}^{\ast}\lr{E+dE,\mu,s,\Psi}\,ds-\int_0^{b\lr{E,\mu,\Psi}}v_{\parallel}^{\ast}\lr{E,\mu,s,\Psi}\,ds\right]\\
=&\lim_{dE\rightarrow 0}\frac{m}{\pi dE}\left[
dE\int_0^b\frac{\p v_{\parallel}^{\ast}}{\p E}\lr{E,\mu,s,\Psi}\,ds+dE\frac{\p b}{\p E}\lr{E,\mu,\Psi}v_{\parallel}^{\ast}\lr{E,\mu,b,\Psi}
\right]\\
=&\frac{1}{\pi}\int_0^b\frac{ds}{v_{\parallel}^\ast\lr{E,\mu,s,\Psi}}=\frac{T_b}{2\pi}=\frac{1}{\omega_b}.
\end{split}
\end{equation}
Observe that here we used the fact that $v_{\parallel}^{\ast}\lr{E,\mu,b,\Psi}=0$. 
It follows that
\begin{equation}
d\Pi=m^2d\ell dv_{\parallel} d\varphi d\Psi d\vartheta_c d\mu=\frac{m\omega_b}{v_{\parallel}^{\ast}}d\ell dJ_{\parallel} d\varphi d\Psi d\vartheta_c d\mu.
\end{equation}
To proceed, it is convenient to approximate the Jacobian $v_{\parallel}^{\ast}/m\omega_b$ with its bounce average,
\begin{equation}
\left\langle\frac{v_{\parallel}^\ast}{m\omega_b}\right\rangle_b=\frac{2}{m\omega_b T_b}\int_0^{t_b}v_{\parallel}^{\ast}\,dt=\frac{b}{m\pi},
\end{equation}
where we used the fact that by definition $\omega_b=\omega_b\lr{E,\mu,\Psi}$ and therefore $\p\omega_b/\p \ell=0$.
The bounce averaged phase space measure thus reads as
\begin{equation}
d\tilde{\Pi}=\frac{m\pi}{b}d\ell dJ_{\parallel} d\varphi d\Psi d\vartheta_c d\mu=m\pi\frac{B}{b}dJ_{\parallel} d\vartheta_c d\mu d^3x.\label{dPiJ}
\end{equation}
The spatial densities $n_p$ and $n_e$ can now be computed as follows. 
For the positron component we have
\begin{equation}
\begin{split}
n_p=&m\pi A_pN_p B\int_0^{\infty}d\mu\int_{0}^{\infty}dJ_{\parallel}\int_0^{2\pi}d\vartheta_c b^{-1}\exp\left\{-\beta_p\lr{\frac{1}{2}\omega_bJ_{\parallel}+\mu B+e\Phi}-\gamma_p\mu-\zeta_p J_{\parallel}\right\}\\
=&2m\pi^2A_p N_pB\int_0^{\infty}d\mu\int_0^{\infty}dJ_{\parallel}b^{-1}\exp\left\{-\beta_p\lr{\frac{1}{2}\omega_bJ_{\parallel}+\mu B+e\Phi}-\gamma_p\mu-\zeta_p J_{\parallel}\right\}.\label{npJ1}
\end{split}
\end{equation}
To simplify \eqref{npJ1} we now follow the approach developed in \cite{YosPTEP}. 
In general, the expression of the bounce frequency $\omega_b$ is a function of $J_{\parallel}$, $\mu$, and $\Psi$. 
However, the dependence on the bounce action $J_{\parallel}$ disappears in the limit $b/L<<1$ in which
the bounce orbit size $b$ is shorter than the characteristic magnetic field line length $L$. 
Indeed, in this limit we may expand the magnetic field in powers of $\ell$ around the
equatorial point $\ell=0$ of the dipole to obtain a second order equation for the bounce position, 
\begin{equation}
E=\mu B\lr{0,\Psi}+\frac{1}{2}\mu\frac{\p^2 B}{\p\ell^2}\lr{0,\Psi}b^2.
\end{equation}
Here, we used the fact that $\p B/\p\ell=0$ at $\ell=0$. 
It is also worth observing that in a dipole field $\p^2 B/\p\ell^2\lr{0,\Psi}>0$ since $B$ has a minimum at $\ell=0$ for each magnetic surface $\Psi$. 
Setting $B_0=B\lr{0,\Psi}$ and $B''_0=\lr{\p^2 B/\p\ell^2}\lr{0,\Psi}$, 
the positive bounce point $b$ can therefore be approximated as
\begin{equation}
b=\sqrt{\frac{2\lr{E-\mu B_0}}{\mu B''_0}}
.\label{bp}
\end{equation}
For a typical particle at $\ell=0$, the energy $E$ can be decomposed into a perpendicular component $E_{\perp 0}=\mu B_0$ 
and a parallel component $E_{\parallel 0}=\alpha_{0p}E_{\perp 0}=\alpha_{0p}\mu B_0$, where $\alpha_{0p}\lr{\Psi}> 0$ is the local positron temperature anisotropy 
at the equator. We may therefore estimate
\begin{equation}
\frac{1}{b}\approx\sqrt{\frac{B''_0}{2\alpha_{0p} B_0}}.\label{bp2}
\end{equation}
Similarly, at second order the bounce frequency becomes
\begin{equation}
\frac{1}{\omega_b}=\frac{1}{\pi}{\int_0^b\frac{ds}{\sqrt{\frac{2}{m}\lr{E-\mu B_0-\frac{1}{2}\mu B''_0s^2}}}}=\frac{\sqrt{m}}{\pi\sqrt{\mu B''_0}}\int_0^{b\sqrt{\frac{\mu B''_0}{2\lr{E-\mu B_0}}}}\frac{dy}{\sqrt{1-y^2}}
=\frac{\sqrt{m}}{2\sqrt{\mu B''_0}}.\label{omb1}
\end{equation}
Notice that in the last passage equation \eqref{bp} was used.
Observe that the bounce frequency is now a function of $\mu$ and $\Psi$ only. 
Recalling equation \eqref{npJ1}, we thus arrive at the following expression for the positron density
\begin{equation}
\begin{split}
n_p= &2m\pi^2A_pN_p B\sqrt{\frac{B''_0}{2\alpha_{0p}B_0}}\exp\left\{-\beta_p e\Phi\right\}\int_0^{\infty}d\mu\int_0^{\infty}dJ_{\parallel}\exp\left\{-\lr{\beta_p B+\gamma_p}\mu-\lr{\beta_p\sqrt{\frac{\mu B''_0}{m}}+\zeta_p}J_{\parallel}\right\},\\
=&2m\pi^2A_pN_p B\sqrt{\frac{B''_0}{2\alpha_{0p}B_0}}\exp\left\{-\beta_p e\Phi\right\}\int_0^{\infty}d\mu
\frac{\exp\left\{-\lr{\beta_p B+\gamma_p}\mu\right\}}{\beta_p\sqrt{\frac{\mu B''_0}{m}}+\zeta_p}
\\=&2m\pi^2A_pN_p \frac{B}{\sqrt{2\alpha_{0p} B_0}}\frac{\exp\left\{-\beta_p e\Phi\right\}}{\beta_p B+\gamma_p}\int_0^{\infty}\frac{e^{-y}\,dy}{
\beta_p\sqrt{\frac{y}{m\lr{\beta_p B+\gamma_p}}}+{\frac{\zeta_p}{\sqrt{B''_0}}}}.\label{npJ2}
\end{split}
\end{equation}
A similar expression can be obtained for the electron density $n_e$ by flipping the sign of the electric charge, and the Poisson equation \eqref{Mx1} for the electrostatic potential $\Phi$ 
becomes
\begin{equation}
\begin{split}
\Delta\Phi=&-\frac{2em\pi^2}{\epsilon_0}\frac{B}{\sqrt{2B_0}}\\&\left[
\frac{A_pN_p\exp\left\{-\beta_pe\Phi\right\}}{\sqrt{\alpha_{0p}}\lr{\beta_pB+\gamma_p}}\int_0^{\infty}\frac{e^{-y}\,dy}{\beta_p\sqrt{\frac{y}{m\lr{\beta_pB+\gamma_p}}}+\frac{\zeta_p}{\sqrt{B''_0}}}
-
\frac{A_eN_e\exp\left\{\beta_ee\Phi\right\}}{\sqrt{\alpha_{0e}}\lr{\beta_eB+\gamma_e}}\int_0^{\infty}\frac{e^{-y}\,dy}{\beta_e
\sqrt{\frac{y}{m\lr{\beta_eB+\gamma_e}}}+\frac{\zeta_e}{\sqrt{B''_0}}}
\right]
\end{split}\label{PhimuJ}
\end{equation}
The last integral in equation \eqref{npJ2} can be written in terms of special functions.
Furthermore, it gives a simple result in the limit $\zeta_p/\beta_p\omega_b<<1$. Indeed, one obtains
\begin{equation}
n_p= \frac{2\pi\lr{m\pi}^{3/2} A_pN_p}{\beta_p}\frac{B\exp\left\{-\beta_p e\Phi\right\}}{\sqrt{2\alpha_{0p} B_0\lr{\beta_p B+\gamma_p}}}
.\label{npJ3}
\end{equation}
Here, it should be noted that although the Lagrange multiplier $\zeta_p$ has been neglected, 
and thus the conservation of the total bounce action $\mc{J}_p$ is not felt by the ensemble,  
the effect of bounce motion on the particle distribution
does not disappear. This is because bounce dynamics is encapsulated in the term $\omega_b J_{\parallel}/2$ of the Hamiltonian $\tilde{H}_p$ appearing in the distribution function $f_p$.
Observing that in the same limit $\zeta_e/\beta_e\omega_b<<1$ the electron density becomes
\begin{equation}
n_e= \frac{2\pi\lr{m\pi}^{3/2} A_eN_e}{\beta_e}\frac{B\exp\left\{\beta_e e\Phi\right\}}{\sqrt{2\alpha_{0e} B_0\lr{\beta_e B+\gamma_e}}},
\end{equation}
the corresponding form of the Poisson equation \eqref{Mx1} for the electrostatic potential $\Phi$ therefore reads
\begin{equation}
\Delta\Phi=-\frac{2\pi e\lr{m\pi}^{3/2}}{\epsilon_0}\frac{B}{\sqrt{2B_0}}\left[\frac{A_pN_p\exp\left\{-\beta_p e\Phi\right\}}{\beta_p\sqrt{\alpha_{0p}\lr{\beta_p B+\gamma_p}}}-\frac{A_eN_e\exp\left\{\beta_e e\Phi\right\}}{\beta_e\sqrt{\alpha_{0e}\lr{\beta_e B+\gamma_e}}}\right].\label{PhimuJ2}
\end{equation}


\section{The effect of the third adiabatic invariant}

It is possible to develop 
an analogous model where the third adiabatic invariant (flux function) $\Psi$ 
is also preserved by single particle dynamics, and enforce conservation of
total magnetic fluxes $\Psi_p=N_p\int_{\Pi}\Psi f_{p}\,d\Pi$ and $\Psi_e=N_e\int_{\Pi}\Psi f_{e}\,d\Pi$ 
in the corresponding entropy principle to obtain the distribution functions
\begin{equation}
f_{p}=A_p\exp\left\{-\beta_p \tilde{H}_p-\gamma_p\mu-\zeta_p J_{\parallel}-\eta_p\Psi\right\},~~~~
f_{e}=A_e\exp\left\{-\beta_e \tilde{H}_e-\gamma_p\mu-\zeta_e J_{\parallel}-\eta_e\Psi\right\},
\end{equation}
where 
$\eta_p,\eta_e$ are Lagrange multipliers associated with $\Psi_p,\Psi_e$. 
It is not difficult to verify that the density $n_p$ of equation \eqref{npJ2} is modified by an
exponential factor $\exp\left\{-\eta_p\Psi\right\}$. A similar modification applies to $n_e$   
so that the Poisson equation \eqref{PhimuJ} for the electrostatic potential $\Phi$ now reads:
\begin{equation}
\begin{split}
\Delta\Phi=&-\frac{2em\pi^2}{\epsilon_0}\frac{B}{\sqrt{2B_0}}\left[
\frac{A_pN_p\exp\left\{-\beta_pe\Phi-\eta_p\Psi\right\}}{\sqrt{\alpha_{0p}}\lr{\beta_pB+\gamma_p}}\int_0^{\infty}\frac{e^{-y}\,dy}{\beta_p\sqrt{\frac{y}{m\lr{\beta_pB+\gamma_p}}}+\frac{\zeta_p}{\sqrt{B''_0}}}\right.\\
&-\left.
\frac{A_eN_e\exp\left\{\beta_ee\Phi-\eta_e\Psi\right\}}{\sqrt{\alpha_{0e}}\lr{\beta_eB+\gamma_e}}\int_0^{\infty}\frac{e^{-y}\,dy}{\beta_e\sqrt{\frac{y}{m\lr{\beta_eB+\gamma_e}}}+\frac{\zeta_e}{\sqrt{B''_0}}}
\right]
\end{split}\label{PhieqPsi}
\end{equation}
However, it should be emphasized that 
for the third adiabatic invariant to be constant, the
time scale $T_f$ of energy fluctuations must be longer than the
characteristic time scale $T_d$ of drift dynamics $\bol{v}_d$ across the magnetic field $\bol{B}$ in the toroidal direction $\varphi$, i.e. 
\begin{equation}
T_f>>T_d=\frac{2\pi}{\omega_d},\label{tftd}
\end{equation}
where $\omega_d$ is the drift frequency. 
Since $\omega_d$ is mainly determined by the $\bol{E}\cp\bol{B}$, curvature, and gradient drift velocities, 
this frequency is usually smaller than the cyclotron and bounce frequencies,
and \eqref{tftd} poses a rather stringent condition on the allowed turbulent spectrum 
of electromagnetic fluctuations. 

Finally, we remark that the contributions to the particle energy coming from the
guiding center drifts are neglected in the guiding center energies $\tilde{H}_{p},\tilde{H}_e$ 
under the assumption that the corresponding kinetic energy is smaller than the other terms. 
In particular, this is true 
if the kinetic energy associated with the $\bol{E}\cp\bol{B}$ drift
$\bol{v}_{\bol{E}}=\bol{E}\cp\bol{B}/B^2$ satisfies the ordering $m\abs{\bol{v}_{\bol{E}}}^2/2<<
\tilde{H}_p,\tilde{H}_e$. 
Such configuration can be achieved for example 
when the electric potential $\Phi$ is a first order contribution in the ratio $\epsilon\sim \rho_c/L$,
where $\rho_c$ is the Larmor radius and $L$ the characteristic scale length of the magnetic field.
For a plasma with inverse temperature $\beta$, this implies $\beta e\abs{\Phi}\sim\epsilon$ as well as
$\beta\frac{m}{2}\abs{\bol{v}_{\bol{E}}}^2\sim\epsilon^2$, and the energies $\tilde{H}_{p},\tilde{H}_e$ are
accurate at first order in $\epsilon$. 
For additional details on the expression of the guiding center Hamiltonian, see \cite{Cary}. 

\section{Numerical computation of spatial densities and electrostatic  potential}
The aim of this section is to numerically study the behavior of the positron-electron plasma
as described by the model developed in the previous sections for different values of the physical parameters involved. 

\begin{figure}[h]
\hspace*{-0cm}\centering
    \includegraphics[scale=0.56]{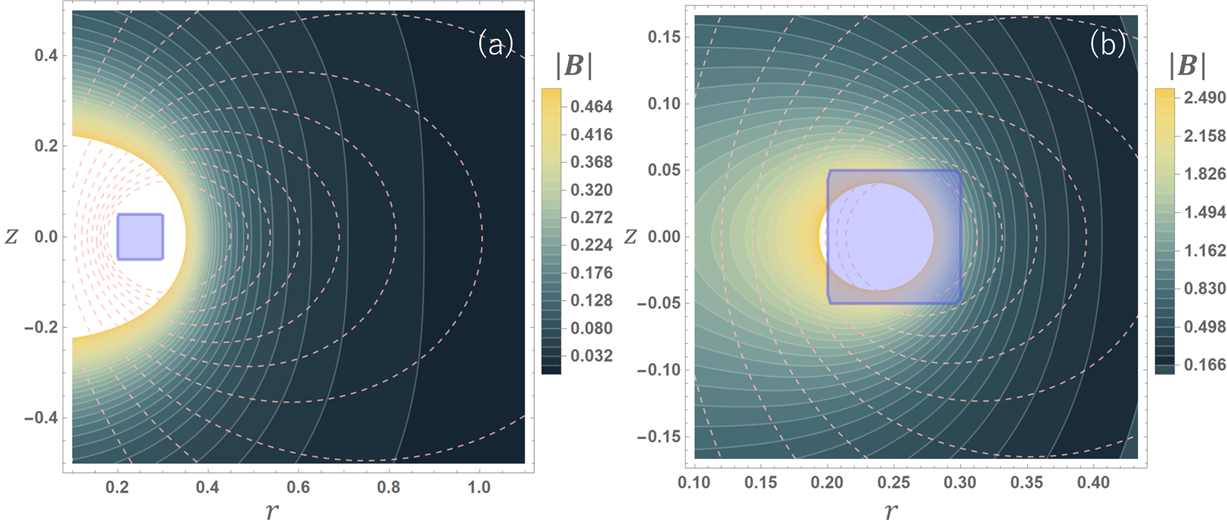}
\caption{\footnotesize (a) Contour plot of the magnetic field strength $\abs{\bol{B}}$ within the domain $\Omega$ enclosed by the positron-electron trap. (b) Contour plot of $\abs{\bol{B}}$ in the region surrounding the current loop. In both (a) and (b) dashed contours correspond to magnetic field lines, while white regions exceed the plotted range of values. The square centered at $r=r_l=0.25\,m$ represents the section of the toroidal box containing the current loop generating the dipole field.} 
\label{fig1}
\end{figure}

For the spatial domain $\Omega$ occupied by the plasma, we consider
a region $r\in\left[r_{0},r_{0}+R\right]$, $z\in[-R/2,R/2]$, $\varphi\in[0,2\pi)$ mimicking an axially symmetric trap with
radial size $R$, height $R$, and a physical axis (center stack) 
whose external boundary is located at $r=r_{0}$.
In particular, we choose $R=1\,m$ and $r_0=0.1\,m$, which are values 
of the same order of the length parameters of the planned positron-electron trap. 
The dipole magnetic field used to confine the plasma is generated by
a current loop of infinitesimal section and radius $r_{l}=0.25\,m$ enclosed in an axially symmetric toroidal box (coil) with squared cross section whose left size is located at $r_{\rm box}=0.2\,m$ (see \cite{Simpson} for the expression of the magnetic field). 
The typical magnetic field generated by the current loop is around $B\sim 0.1 T$ 
and reaches $B\sim 2T$ in close proximity of the bounding box. 
It is also convenient to introduce the characteristic magnetic field $B^{\ast}=\mu_0I/2r_{l}=1.25\,T$ where $\mu_0$ is the vacuum permeability and $I$ the electric current flowing within the current loop. 
The setting described above is shown in figure \ref{fig1}. 

We will now examine the plasma equilibria corresponding 
to the statistical ensembles constructed in sections 2, 3, and 4 separately.

\subsection{$\mu$ Equilibrium}
Consider the positron-electron plasma equilibrium \eqref{fpfe} 
arising from conservation of total magnetic moments $M_p,M_e$.
We shall refer to such equilibrium as a $\mu$ equilibrium. 
The condition for a $\mu$ equilibrium to hold is that the time scale
of the electromagnetic fluctuations driving the system toward the relaxed state 
is longer than the time scale of cyclotron dynamics as described by equation \eqref{tftc}. 
In practice, this means that the term $q\Phi$ occurring within the Hamiltonians $H_p,H_e$ 
evolves over long time scales compared to the cyclotron time scale.   
This ensures that the constancy of the first adiabatic invariant $\mu$ is not broken.   

For the system under consideration, the cyclotron frequency $\omega_c$ is
\begin{equation}
\omega_c=\frac{eB}{m}\approx 10^{10}\,Hz,
\end{equation}
where the typical value $B\sim 0.1\,T$ has been used. 
Denoting with $\omega_f=2\pi/T_f$ the frequency of electromagnetic fluctuations, 
the condition \eqref{tftc} can therefore be written as
\begin{equation}
\omega_f<<\omega_c\approx 10^{10}\,Hz.\label{tftc2}
\end{equation}
It is worth observing that
for a comparable trap 
such as the RT-1 device \cite{YosPRL}, 
the spectrum of electromagnetic fluctuations is well below the $10^5\,Hz$ range. 
We therefore expect the condition \eqref{tftc2} to be  easily fulfilled by the system under examination. We shall therefore assume that 
\eqref{tftc2} holds, and apply the 
equilibrium model developed in section 2. 

In order to evaluate the densities \eqref{npne2}, 
the Poisson equation \eqref{PhiEq} must be solved for the
electrostatic potential $\Phi$.
The physical parameters $A_p$, $A_e$, $N_p$, $N_e$, $\beta_p$, $\beta_e$, $\gamma_p$, and $\gamma_e$ appearing on the right-hand side of equation \eqref{PhiEq} are determined as follows. 
First, the target temperature $T$ for the positron-electron plasma is $T\approx T_p\approx T_e\approx 10\,eV$, where $T_p$ and $T_e$ denote the temperatures of positrons and electrons respectively. 
This fixes the inverse temperatures $\beta_p^{-1}=k_B T_p$ and $\beta_e^{-1}=k_BT_e$, with $k_B$ the Boltzmann constant. The target spatial density $n$ is $n\approx n_p\approx n_e\approx 10^{11}\,m^{-3}$ up to $10^{12}\,m^{-3}$. Since the volume of the positron-electron plasma trap is of the order $\Omega\approx 1\,m^3$, 
we consider a total particle number of the order $N\approx N_p\approx N_e\approx 10^{11}$. Next, an estimate of the chemical potentials $\gamma_p$ and $\gamma_e$  
can be obtained by observing that the changes in the total energies $E_p,E_e$ 
caused by the addition of 
magnetic moments $dM_p,dM_e$ to the system are given by $\beta_p dE_p=\gamma_p dM_p$ and $\beta_e dE_e=\gamma_e dM_e$. 
The order of magnitude of $\gamma_p,\gamma_e$ is therefore expected to be $\gamma_p\approx\beta_p H_p/\mu\approx \beta_p B$ and 
$\gamma_e\approx\beta_e H_e/\mu\approx \beta_e B$. 
Finally, the parameters $A_p,A_e$, which represent measures of the volume $\Pi$ 
occupied by the plasma in the phase space, can be estimated 
by observing that for a Maxwell-Boltzmann distribution
\begin{equation}
f_B\lr{\bol{x},\bol{p}}=A\exp\left\{-\beta\frac{p^2}{2m} \right\},
\end{equation}
one has
\begin{equation}
A=\frac{1}{\int_{\Pi}\exp\left\{-\beta \frac{p^2}{2m}\right\}\,d^3x d^3p}=\frac{1}{\Omega}\lr{\frac{\beta }{2\pi m}}^{\frac{3}{2}}.
\end{equation}
We therefore expect that $A_p\approx \Omega^{-1}\lr{\beta_p /2\pi m}^{3/2}$ and $A_e\approx \Omega^{-1}\lr{\beta_e /2\pi m}^{3/2}$.

The last ingredient needed to numerically solve equation 
\eqref{PhiEq} is the value of the potential $\Phi$ on the
boundary $\p\Omega$. Note that the boundary $\p\Omega$ consists of the vessel boundary $\p\Omega_{\rm v}$ and the coil surface $\p\Omega_{c}$. 
At the vessel boundary $\p\Omega_{\rm v}$ the electrostatic potential $\Phi$ 
is grounded, i.e. $\Phi=0$ on $\p\Omega_{\rm v}$.
The coil generating the dipole magnetic field is levitated  
without mechanical contact through a secondary magnet located in the upper region of the trap. Therefore, the value 
$\Phi_c$ of the electrostatic potential on the coil surface $\p\Omega_c$ is determined 
by the amount of charged particles that hit it. 
In particular, electrons penetrating the coil surface 
push $\Phi_c$ toward negative values, while positrons
tend to increase $\Phi_c$ (note that even if positrons annihilate with electrons after reaching the coil surface, this results in a positive increase of the coil charge). 
An upper bound to $\Phi_c$ can be obtained from the Poisson equation \eqref{Mx1} through the scaling  
\begin{equation}
\Phi\approx \frac{e n R^2}{\epsilon_0}\approx 10^3\,V,\label{PhiEst}
\end{equation}
where $R$ is the radial size of the device and the value $n\approx 10^{11}\,m^{-3}$ has been used. 
In practice $\Phi_c$ is expected to be much smaller than \eqref{PhiEst} since 
the charge separation $e\lr{n_p-n_e}$ 
will be sensibly smaller than the characteristic charge density $en\approx 10^{-8}\,C m^{-3}$ of each individual species. 
In summary, the boundary conditions that will be used to evaluate the Poisson equation \eqref{PhiEq} are
\begin{equation}
\Phi=0~~~~{\rm on}~~\p\Omega_{\rm v},~~~~\Phi=\Phi_c~~~~{\rm on}~~\p\Omega_c,\label{BC}
\end{equation}
with $\abs{\Phi_c}<10^3\,V$.

\begin{figure}[h!]
\hspace*{-0cm}\centering
    \includegraphics[scale=0.82]{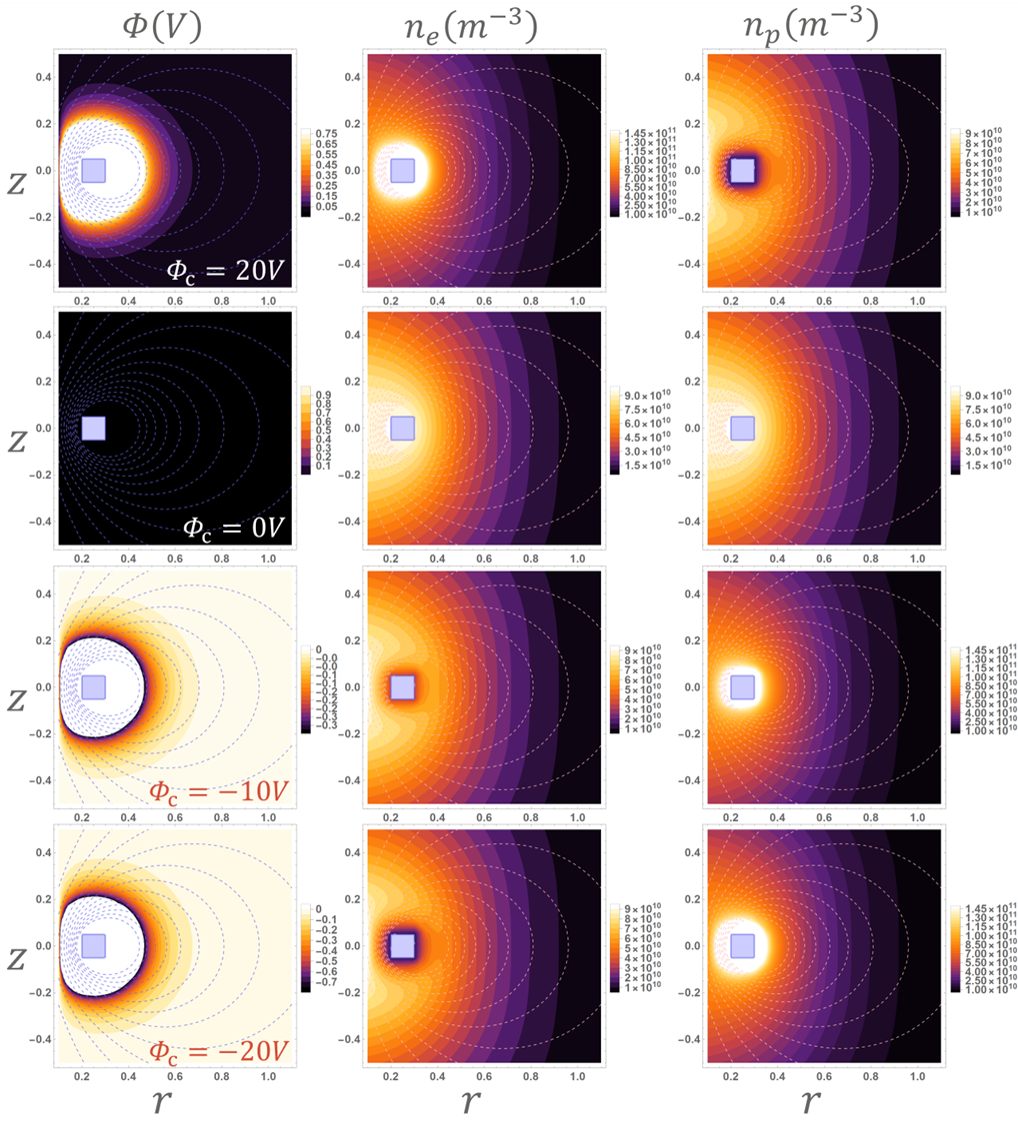}
\caption{\footnotesize Contour plots of electrostatic potential $\Phi$ (left column), electron density $n_e$ (center column), and positron density $n_p$ (right column) obtained by numerical solution of \eqref{PhiEq} with boundary conditions \eqref{BC} in the $\lr{r,z}$ plane. Each row corresponds to a different value of the coil potential $\Phi_c$:  from top row to bottom row the values are $20\,V$, $0\,V$, $-10\,V$, and $-20\,V$.
The physical parameters used in this simulation are $A_p=A_e=3.8\,\Omega^{-1}\lr{\beta /2\pi m}^{3/2}$ with $\beta^{-1}=k_BT$ and $T=T_p=T_e=10\,eV$, $N_p=N_e=10^{11}$, and $\gamma_p=\gamma_e=0.1\beta B^{\ast}$ with $B^{\ast}=1.25\,T$ the  characteristic magnetic field. Dashed contours represent magnetic field lines. White regions exceed the plotted range of values.}
\label{fig2}
\end{figure}

Figure \ref{fig2} shows the
modification of spatial densities $n_p,n_e$ and electrostatic potential $\Phi$ obtained from numerical solution of \eqref{PhiEq} when varying the coil potential $\Phi_c$ while keeping the other physical parameters fixed. 
As one may expect, a negatively charged coil attracts the positron component while repelling electrons. The converse occurs for a positively charged coil. 
The other essential feature of figure \ref{fig2} is the effect of the first adiabatic invariant $\mu$ on the macroscopic equilibrium state of the system: both positrons and electrons tend to accumulate in regions of higher magnetic field strength. This feature combined with the dependence with respect to the coil potential implies that when  $\Phi_c> 0$ positrons   preferentially populate the interior region close to the center stack and enclosed by the coil, $r<r_{\rm box}$, while electrons   surround the coil surface (first row in figure \ref{fig2}). 
The opposite occurs when $\Phi_c>0$ (third and fourth rows in figure \ref{fig2}). 
Inspection of \eqref{PhiEq}
also suggests that 
if the thermodynamic parameters of the two species are equal, 
i.e. $A_p=A_e$, $N_p=N_e$,  $\beta_p=\beta_e$, and $\gamma_p=\gamma_e$, and $\Phi_c=0$ on $\p\Omega_c$, then $\Phi$ vanishes throughout $\Omega$ while the spatial densities are identical, $n_p=n_e$. This case is shown 
in the second row of figure \ref{fig2}. 

\begin{figure}[h!]
\hspace*{-0cm}\centering
    \includegraphics[scale=1]{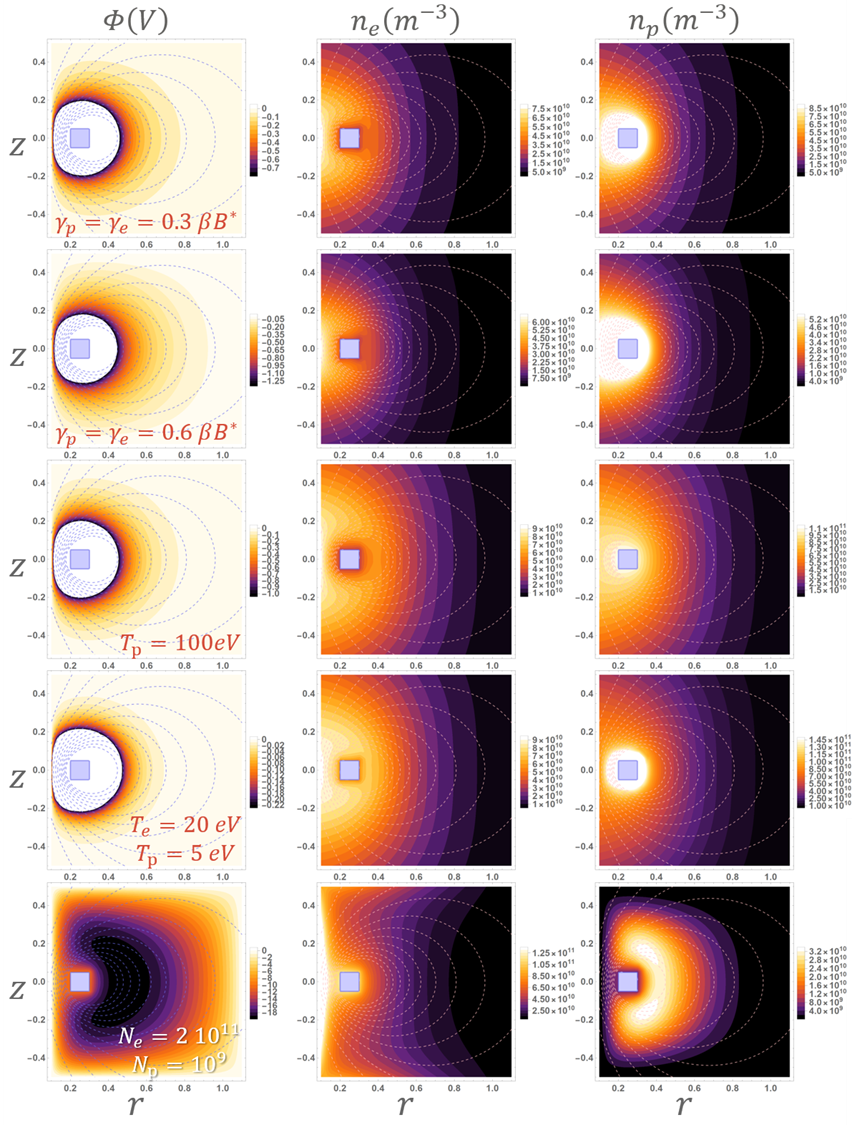}
\caption{\footnotesize Contour plots of electrostatic potential $\Phi$ (left column), electron density $n_e$ (center column), and positron density $n_p$ (right column) obtained by numerical solution of \eqref{PhiEq} with boundary conditions \eqref{BC} in the $\lr{r,z}$ plane. Each row corresponds to a modification of one of the physical parameters used for the case shown in the third row of figure \ref{fig2}. 
In the first and second rows the chemical potentials have been increased to $\gamma_p=\gamma_e=0.3\beta B^{\ast}$ and $\gamma_p=\gamma_e=0.6\beta B^{\ast}$ respectively. In the third and fourth row the temperatures have been changed to $T_p=100\,eV$ and $T_e=4T_p=20\,eV$ respectively. In the fifth row the particles numbers have been changed to $N_e=2 \cdot 10^{11}$ and $N_p=10^9$. 
The relevant parameter changes have also been highlighted within each $\Phi$ plot. 
Dashed contours represent magnetic field lines. White regions exceed the plotted range of values.}
\label{fig3}
\end{figure}

Taking the third row in figure \ref{fig2} as reference case, figure \ref{fig3} summarizes how spatial densities $n_p,n_e$ and electrostatic potential $\Phi$ change when physical parameters are modified. 
While equilibrium profiles are not too sensitive to changes in the chemical potentials $\gamma_p,\gamma_e$ and in the
inverse temperatures $\beta_p,\beta_e$, 
abrupt changes occur when 
there is asymmetry in the number of protons and electrons. The bottom row of figure \ref{fig3} shows the case  $N_e=2\cdot 10^{11}>>N_p=10^9$. 
Note that positrons are almost completely expelled from the interior region $r<r_{\rm box}$ and form a radiation belt like structure on the exterior side of the coil, while electrons are mostly found in proximity of the center stack located at $r=r_0$.


\subsection{$\mu$-$J_{\parallel}$ Equilibrium}

We now consider the case in which individual particles preserve the second adiabatic invariant $J_{\parallel}$. 
For the bounce action to be constant, the time scale $T_f$ of electromagnetic fluctuations
must be longer than the period $T_b$ of a bounce oscillation as
described by equation \eqref{tftb}. 
An estimate of the bounce frequency $\omega_b$ (and thus of $T_b$) can be obtained with the aid of equation \eqref{omb1}, which rests on the hypothesis $\mu B>>e\abs{\Phi}$ or $\beta_p^{-1},\beta_e^{-1}>>e\abs{\Phi}$.
Therefore, assuming these conditions to hold, the typical bounce frequency is
\begin{equation}
\omega_b=2\sqrt{\frac{\mu B_0''}{m}}\approx 2\frac{\sqrt{\beta^{-1}}}{R\sqrt{m}}\approx 10^6\,Hz.
\end{equation}
Here, the values $\beta^{-1}=k_BT$ with $T=10\,eV$ and the trap scale length $R=1\, m$ have been used. 
Hence, the condition \eqref{tftb} now reads
\begin{equation}
\omega_f<<\omega_b\approx 10^6\,Hz.\label{tftb2}
\end{equation}
Notice that since $\omega_b<<\omega_c$ (recall equation \eqref{tftc2}) 
if the condition \eqref{tftb2}
is satisfied then the first adiabatic invariant $\mu$ is 
also preserved. Again, 
experimental evidence from the RT-1 device suggests that 
the requirement \eqref{tftb2} is usually fulfilled since $\omega_f<10^5\, Hz$ there  \cite{YosPRL}. 

In addition to the physical parameters $A_p$, $A_e$, $N_p$, $N_e$, $\beta_p$, $\beta_e$, $\gamma_p$, and $\gamma_e$ 
already encountered for the $\mu$ equilibrium, 
the $\mu$-$J_{\parallel}$ equilibrium described by the Poisson equation \eqref{PhimuJ}
includes the temperature anisotropies $\alpha_{0p}$ and $\alpha_{0e}$ and the Lagrange multipliers $\zeta_p$ and $\zeta_e$ associated with conservation of the total bounce actions $\mc{J}_p$ and $\mc{J}_e$. In the following, we study a plasma with $\alpha_{0p}=\alpha_{0e}=1$. Notice that if the plasma is isotropic  $\alpha_{0p}=\alpha_{0e}=1/2$ since perpendicular motion carries twice the degrees of freedom of parallel dynamics. Nevertheless, as long as $\alpha_{0p}$ and $\alpha_{0e}$ are constant, the spatial densities profiles remain unchanged, but they are only scaled by factors $1/\sqrt{\alpha_{0p}}$ and $1/\sqrt{\alpha_{0e}}$ respectively (recall equation \eqref{npJ2}). 
We shall also consider the limiting case $\zeta_p/\beta_p\omega_b<<1$ and $\zeta_e/\beta_e\omega_b<<1$ to simplify the evaluation of the integrals on the right-hand side of \eqref{PhimuJ}. Then, the relevant Poisson equation is given by \eqref{PhimuJ2} where $\zeta_p,\zeta_e$ do not appear explicitly. Notice that physically this implies that the total bounce actions $\mc{J}_p,\mc{J}_e$ are not preserved during the 
relaxation 
of the system. However, the effect of bounce dynamics is still felt by the ensemble 
through the term $\omega_bJ_{\parallel}/2$ appearing in the energies $\tilde{H}_p,\tilde{H}_e$ and the bounce averaged phase space measure $d\tilde{\Pi}$ of \eqref{dPiJ}. 

\begin{figure}[h!]
\hspace*{-0cm}\centering
    \includegraphics[scale=0.86]{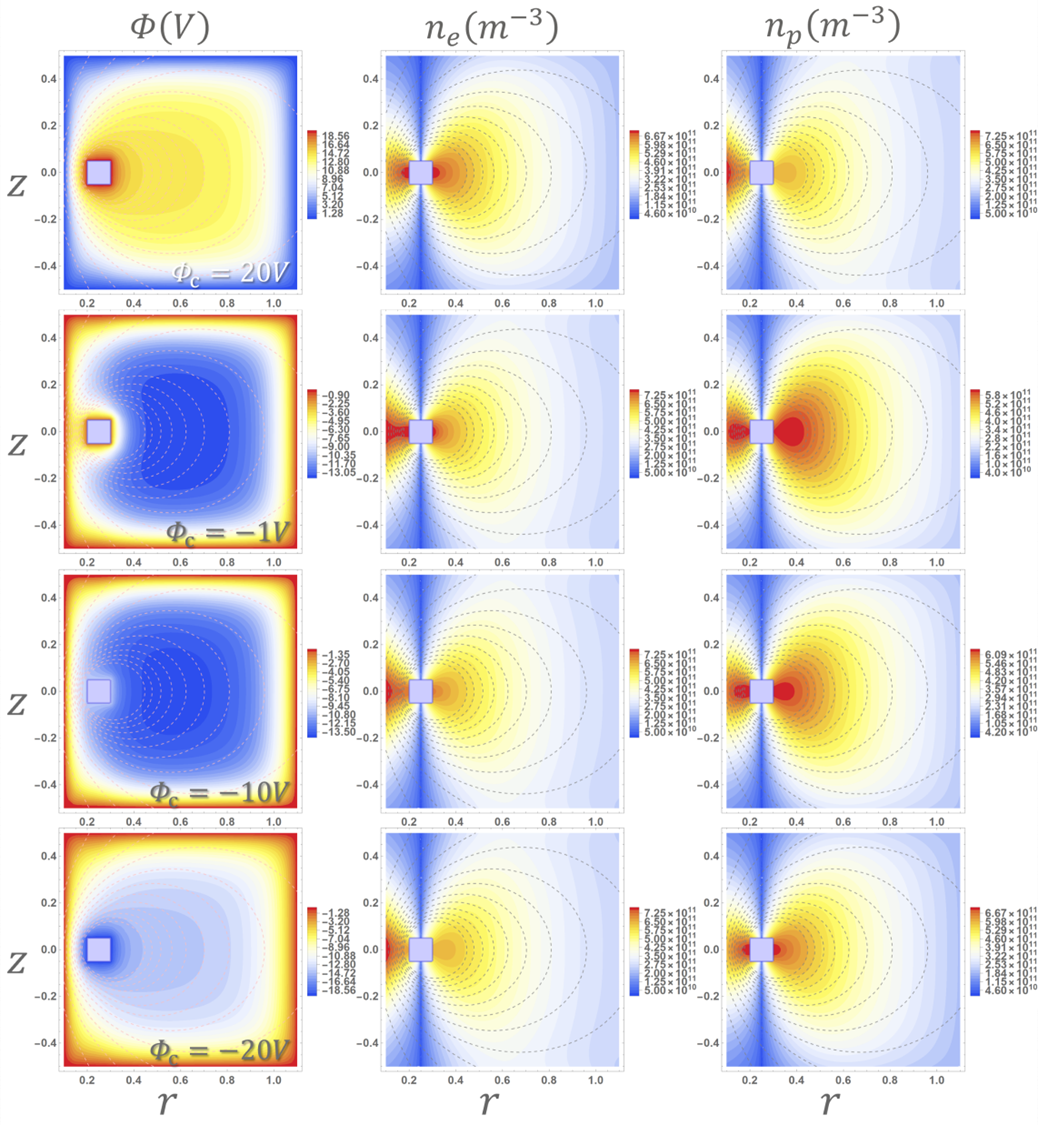}
\caption{\footnotesize Contour plots of electrostatic potential $\Phi$ (left column), electron density $n_e$ (center column), and positron density $n_p$ (right column) obtained by numerical solution of \eqref{PhimuJ2} with boundary conditions \eqref{BC} in the $\lr{r,z}$ plane. Each row corresponds to a different value of the coil potential $\Phi_c$:  from top row to bottom row the values are $20\,V$, $-1\,V$, $-10\,V$, and $-20\,V$.
The physical parameters used in this simulation are $A_p=A_e=2.4\,\Omega^{-1}\lr{\beta /2\pi m}^{3/2}$ with $\beta^{-1}=k_BT$ and $T=T_p=T_e=100\,eV$, $N_p=6\cdot10^{11}$, $N_e=8\cdot10^{11}$, $\alpha_{0p}=\alpha_{0e}=1$, and $\gamma_p=\gamma_e=0.05\beta B^{\ast}$ with $B^{\ast}=1.25\,T$ the  characteristic magnetic field. Dashed contours represent magnetic field lines.}
\label{fig4}
\end{figure}

Figure \ref{fig4} shows $\mu$-$J_{\parallel}$ equilibria obtained by numerical solution of the Poisson equation \eqref{PhimuJ2}
for different values of the coil potential $\Phi_c$ when the other physical parameters are kept constant. 
As in the case of $\mu$ equilibria, a charged coil tends to push the species with charge of the same sign  
in the internal region $r<r_{\rm box}$ while attracting the species with opposite charge. 
Next, observe that as in the previous case the conservation of the magnetic moment $\mu$ 
results in the tendency of particles to accumulate in regions of high magnetic field strength. 
However, there is a peculiar feature of $\mu$-$J_{\parallel}$ equilibria : 
bounce dynamics squeezes spatial density profiles along the equatorial line $z=0$.    
As a result, radiation belt like structures are formed on both sides of the coil. 
This behavior is a consequence of the term $1/\sqrt{B_0}$ appearing in the expression of the 
spatial densities $n_p,n_e$ (recall equation \eqref{npJ3}), which is related to the characteristic bounce velocity $v_{b}$ given by
\begin{equation}
v_b=\frac{2b}{T_b}=\frac{1}{\pi}b\omega_b=\sqrt{\frac{2\alpha_pB_0\mu}{m}},
\end{equation}
where equations \eqref{bp2} and \eqref{omb1} were used. 
Since $n_p,n_e\propto 1/\sqrt{B_0}$, and the characteristic bounce velocity $v_b\propto \sqrt{B_0}$ 
is higher in regions of stronger magnetic field strength $B_0=B\lr{r,0}$, the spatial densities
have a minimum in correspondence of the maximum of $B_0$, which is located at the radial position
of the current loop $r=r_{l}$. 
Physically, the formation of radiation belt like structures can be understood as follows. 
Particles with a high characteristic bounce velocity are statistically less likely to occur, 
and the net result is balance between the tendency of particles to populate
regions of high magnetic field strength $B$ as a result of the first adiabatic invariant,  
and the depletion effect caused by the second adiabatic invariant at those radial positions where $B_0$, and thus $v_b$, are higher. 

\begin{figure}[h!]
\hspace*{-0cm}\centering
    \includegraphics[scale=1.02]{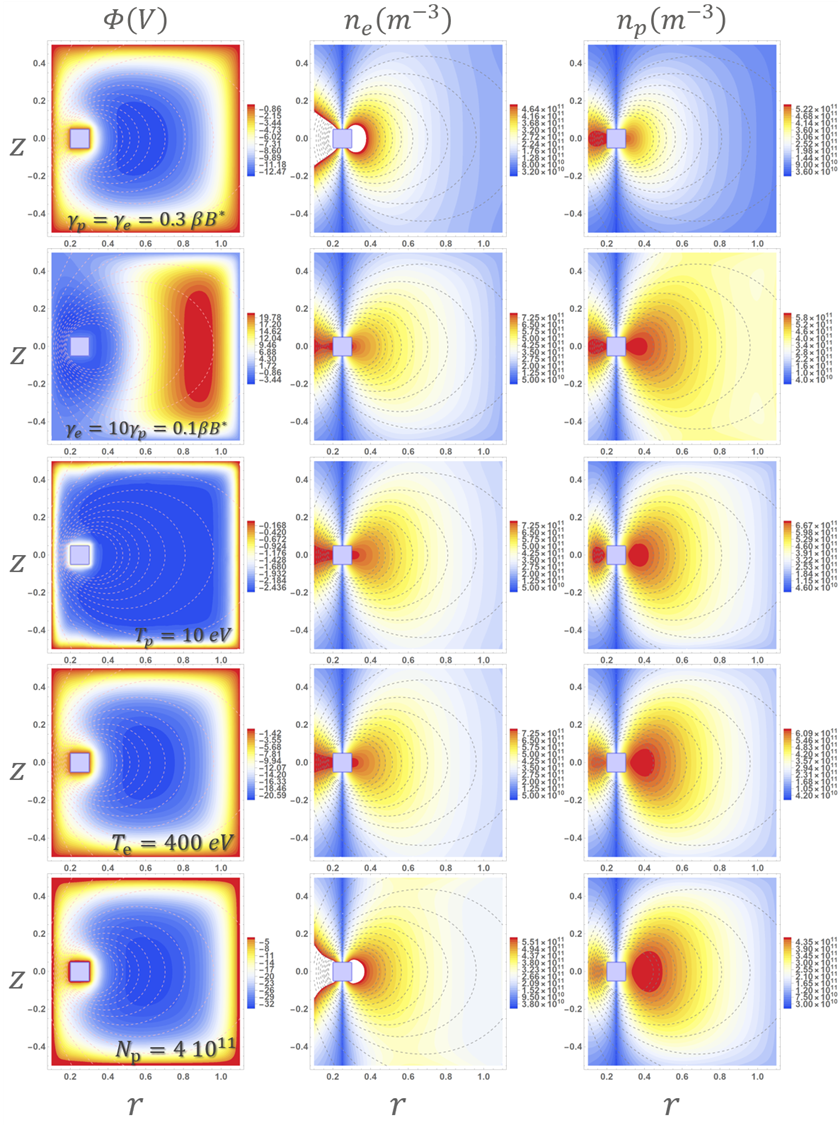}
\caption{\footnotesize Contour plots of electrostatic potential $\Phi$ (left column), electron density $n_e$ (center column), and positron density $n_p$ (right column) obtained by numerical solution of \eqref{PhimuJ2} with boundary conditions \eqref{BC} in the $\lr{r,z}$ plane. Each row corresponds to a modification of one of the physical parameters used for the case shown in the second row of figure \ref{fig4}. 
In the first and second rows the chemical potentials have been modified to $\gamma_p=\gamma_e=0.3\beta B^{\ast}$ and $\gamma_e=10\gamma_p=0.1\beta B^{\ast}$ respectively. In the third and fourth row the temperatures have been changed to $T_p=10\,eV$ and $T_e=400\,eV$ respectively. In the fifth row the positron number has been changed to $N_p=4 \cdot 10^{11}$. 
The relevant parameter changes have also been highlighted within each $\Phi$ plot. 
Dashed contours represent magnetic field lines. White regions exceed the plotted range of values.}
\label{fig5}
\end{figure}

The dependence of spatial densities $n_p,n_e$ and electrostatic potential $\Phi$ on the various physical parameters is shown in figure \ref{fig5}. 
Each row corresponds to a variation of one of the parameters used in the example plotted in  the second row of figure \ref{fig4}. 

\subsection{$\mu$-$J_{\parallel}$-$\Psi$ Equilibrium}
If the frequency of electromagnetic fluctuations is sufficiently small, 
the third adiabatic invariant $\Psi$ is preserved in addition to $\mu$ and $J_{\parallel}$. 
For the system under examination, the drift frequency is determined by the 
frequency of the toroidal rotation around the vertical axis. 
At equilibrium, the toroidal drift velocity $v_{\varphi}=r\bol{v}_d\cdot\nabla\varphi$ is given by the toroidal component of the guiding center drift velocity $\bol{v}_d=\bol{v}_{\bol{E}}+\bol{v}_{\bol{k}}$,
which is the sum of $\bol{E}\cp\bol{B}$ drift
\begin{equation}
\bol{v}_{\bol{E}}=\frac{\bol{E}\cp\bol{B}}{B^2}=\frac{\p_{\ell}\cp\nabla\Phi}{B},\label{vE}
\end{equation}
and gradient plus curvature drift
\begin{equation}
\bol{v}_{\bol{k}}=\frac{2E_{\parallel}+E_{\perp}}{qB}\p_{\ell}\cp\p_{\ell}^2=
\frac{1+2\alpha}{q\beta B}\p_{\ell}\cp\p_{\ell}^2,\label{vk}
\end{equation}
where $\p_{\ell}^2=\p_{\ell}\cdot\nabla\p_{\ell}$ is the curvature of the magnetic field, 
and $\beta$ and $\alpha=E_{\parallel}/E_{\perp}$ are the inverse temperature and the
temperature anisotropy of the particle species under consideration. 
In deriving \eqref{vk} we used the fact that in a vacuum magnetic field the term  $\p_{\ell}\cp\nabla B$ and the curvature $\p_{\ell}^2$ are related by 
$\p_{\ell}\cp\lr{\nabla\cp\bol{B}}=\p_{\ell}\cp\lr{\nabla B\cp\p_{\ell}}-B\p_{\ell}^2=\bol{0}$, which implies $\p_{\ell}\cp\nabla B=B\p_{\ell}\cp\p_{\ell}^2$. 
When the plasma temperature $\beta^{-1}$ and the magnetic field curvature $\p_{\ell}^2$ 
are sufficiently small, the $\bol{E}\cp\bol{B}$ 
drift velocity becomes the dominant contribution to $\bol{v}_d$. Then, the order of the 
drift frequency can be evaluated as 
\begin{equation}
\omega_d\approx \frac{E}{B}R\approx \frac{\Phi}{B}\approx 10^4\, Hz,\label{omegad}
\end{equation}
where the estimate  \eqref{PhiEst} for the electrostatic potential $\Phi$, the typical magnetic field $B\sim 0.1\,T$, and the trap radial size $R=1\,m$ were used. 
Unfortunately, experimental evidence \cite{YosPRL} suggests that the frequency of  
electromagnetic fluctuations within the trap will be comparable to \eqref{omegad}, 
implying that the criterion for the existence of a third adiabatic invariant $\Psi$,
\begin{equation}
\omega_f<<\omega_d,
\end{equation}
will not be satisfied. 
Nevertheless, it is instructive to explore how the 
presence of the third adiabatic invariant $\Psi$ modifies the spatial density profiles 
and the electrostatic potential at equilibrium.

\begin{figure}[h]
\hspace*{-0cm}\centering
    \includegraphics[scale=0.72]{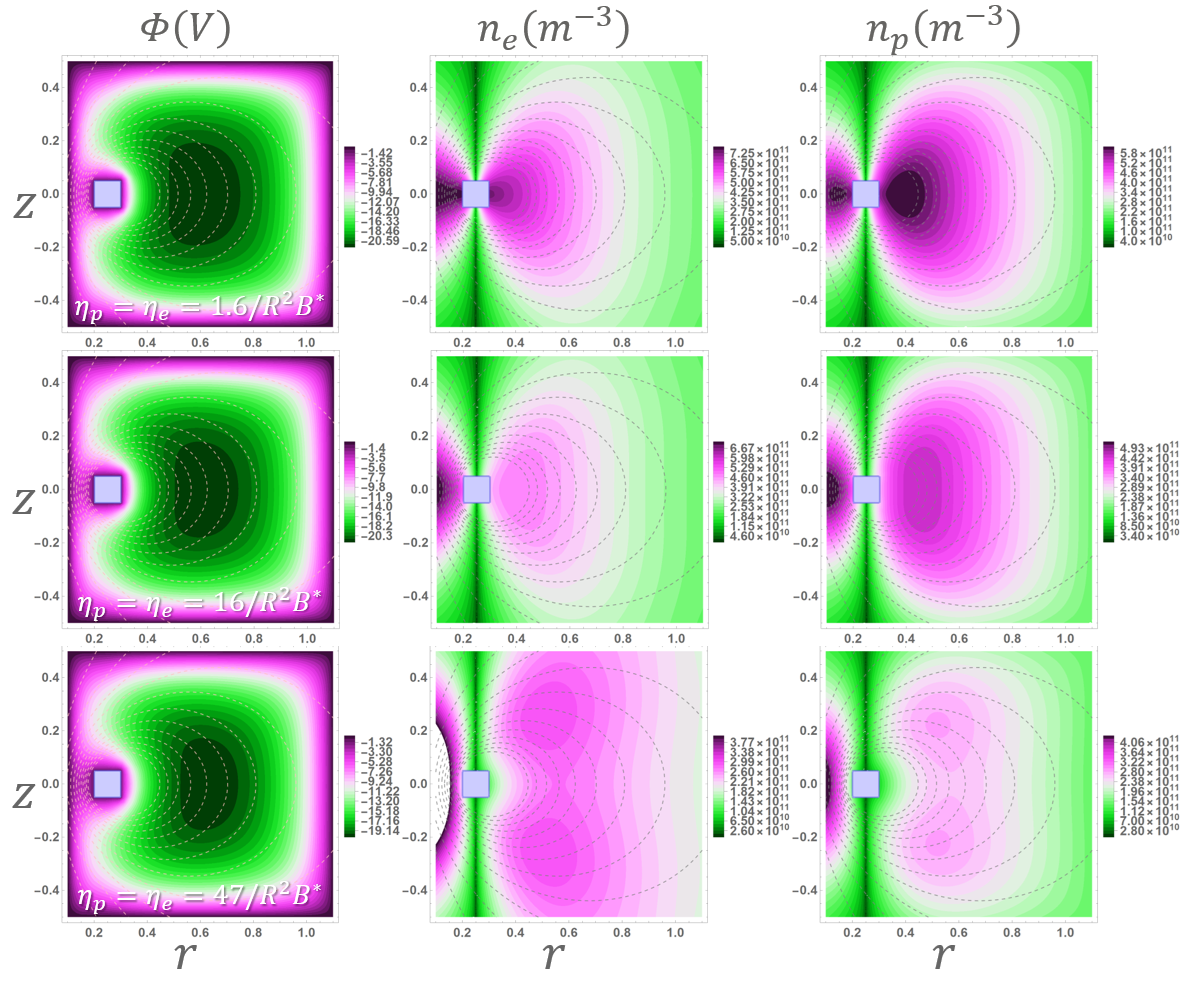}
\caption{\footnotesize Contour plots of electrostatic potential $\Phi$ (left column), electron density $n_e$ (center column), and positron density $n_p$ (right column) obtained by numerical solution of \eqref{PhieqPsi} with boundary conditions \eqref{BC} in the $\lr{r,z}$ plane. Each row corresponds to a modification of one of the physical parameters used for the case shown in the second row of figure \ref{fig4}. 
In the first row $\eta_p=\eta_e=1.6/R^2B^{\ast}$. In the second row $\eta_p=\eta_e=16 /R^2B^{\ast}$. 
In the third row $\eta_p=\eta_e=47/ R^2B^{\ast}$.
Here $R=1 m$ is the radial size of the trap while $B^{\ast}=1.25 T$ the characteristic magnetic field. 
The relevant parameter changes have also been highlighted within each $\Phi$ plot. 
Dashed contours represent magnetic field lines. White regions exceed the plotted range of values.}
\label{fig6}
\end{figure}

Figure \ref{fig6} shows the spatial densities $n_p,n_e$ and the electrostatic potential $\Phi$ 
obtained by numerical solution of \eqref{PhieqPsi} for different values of the chemical potentials $\eta_p,\eta_e$ associated with the third adiabatic invariant $\Psi$. 
The effect of the preservation of total magnetic fluxes $\Psi_p,\Psi_e$ 
is felt by the system only for large values of $\eta_p,\eta_e$, corresponding to large changes in energy $\beta_p dE_p=\eta_pd\Psi_{p}\approx 10$, $\beta_e dE_e=\eta_ed\Psi_{e}\approx 10$ when magnetic fluxes 
$d\Psi_p,d\Psi_e\approx R^2 B^\ast\approx 1\,m^{2}T$ 
are added to the system.   
The spatial densities are initially flattened at the equator (compare the spatial densities of first and second rows in figure \ref{fig6}), eventually splitting
into upper and lower lobes separated by the equatorial line $z=0$ (compare the spatial densities of first and third rows in figure \ref{fig6}). 
Physically, this behavior can be understood as follows. 
The kinetic energy associated with toroidal drift motion can be written as
\begin{equation}
K_{\varphi}=\frac{1}{2}mv_{\varphi}^2=\frac{1}{2mr^2}\lr{p_{\varphi}-q\Psi}^2. 
\end{equation}
This term does not appear in the guiding center Hamiltonians \eqref{Hgc} because it is usually smaller than the kinetic energies associated with parallel and cyclotron dynamics. In particular, this implies that $p_{\varphi}\approx q\Psi$, with $p_{\varphi}$ the canonical momentum of a charged particle in the $\varphi$ direction and where we used the fact that in a dipole magnetic field the vector potential is $\bol{A}=\Psi\nabla\varphi$. Since states with large 
deviations in the 
canonical momentum $p_{\varphi}\approx q\Psi$ 
would break the conservation of the total magnetic flux $\Psi_{\rm tot}$, 
regions with large $\Psi$ 
are penalized in 
the distribution functions through the factors $e^{-\eta_p\Psi}$ and $e^{-\eta_e\Psi}$. 
If the chemical potentials $\eta_p,\eta_e$ are sufficiently large, 
this effect prevails on the tendency caused by the first adiabatic invariant $\mu$ to concentrate particles in regions with strong magnetic field strength $B$, resulting in a preferential depletion of the equatorial region outside the coil (recall that along magnetic field lines, which correspond to level sets of $\Psi$, the magnetic field strength increases when approaching the coil from the outside).



\section{Toroidal Rotation}

In this section we consider basic aspects pertaining to the macroscopic toroidal rotation properties of the positron-electron plasma. Using equations \eqref{vE} and \eqref{vk} and recalling that $\bol{B}=\nabla\Psi\cp\nabla\varphi$,  
the toroidal drift velocity has expression
\begin{equation}
\begin{split}
v_{\varphi} 
=&\frac{r}{\abs{\nabla\Psi}^2}\lr{\nabla\Phi+\frac{1+2\alpha}{q\beta}\p_{\ell}^2}\cdot\nabla\Psi\\
=&r\frac{\p\Phi}{\p\Psi}+\frac{r}{\abs{\nabla\Psi}^2}\lr{\frac{\p\Phi}{\p\ell}\nabla\ell+\frac{1+2\alpha}{q\beta}\p_{\ell}^2}\cdot\nabla\Psi.\label{vphi}
\end{split}
\end{equation}
It is well known that a single species plasma trapped in 
a straight homogeneous magnetic field $\bol{B}=B_{z}\nabla z$, $B_z\in\mathbb{R}$ 
relaxes to a self-organized rigidly rotating equilibrium with
toroidal rotation velocity $v_{\varphi}=\omega_z r$, $\omega_z\in\mathbb{R}$ (see e.g. \cite{Dubin}). 
This result can be recovered from \eqref{vphi} as follows. 
First, observe that a straight magnetic field can be expressed as $\bol{B}=\nabla\Psi\cp\nabla\varphi$ with $\Psi=B_zr^2/2+c_{\Psi}$ and $c_{\Psi}\in\mathbb{R}$. Furthermore, since $\ell=z$ and $\p^2_{\ell}=\bol{0}$, the toroidal drift velocity reduces to
\begin{equation}
v_{\varphi}=r\frac{\p\Phi}{\p\Psi}.\label{vphistraight}
\end{equation}
In addition, the homogeneity of the magnetic field implies that bounce motion is absent, 
while the conservation of the first adiabatic invariant $\mu$ does not affect the 
profile of the spatial density (recall equation \eqref{npne2}).
The only relevant constraint is thus that given by the third adiabatic invariant $\Psi$.  
Indeed, denoting with $\bol{v}=\bol{v}_{\parallel}+\bol{v}_{\bol{E}}$ the velocity of a charged particle (other drifts are absent due to the homogeneity of the magnetic field) one has  
\begin{equation}
\frac{d\Psi}{dt}=\bol{v}\cdot\nabla\Psi=-\frac{\p\Phi}{\p\varphi}=0,
\end{equation}
where we used the fact that 
the electrostatic potential is expected to be axisymmetric, $\p\Phi/\p\varphi=0$. 
The conservation of 
the total magnetic flux $\Psi_{\rm tot}=N\int_{\Pi}\Psi f\,d\Pi$, with $f$ the particle distribution function and $N$ the particle number, then leads to a spatial density distribution
\begin{equation}
n=A\exp\left\{-\beta q\Phi-\eta\Psi\right\},
\end{equation}
where $A$ is a positive real constant and $\eta$ a Lagrange multiplier associated with  preservation of total magnetic flux. 
The Poisson equation for the electrostatic potential $\Phi$
in an infinite vertically symmetric $\p\Phi/\p z=0$ plasma column
now reads 
\begin{equation}
\frac{1}{r}\frac{\p}{\p r}\lr{r\frac{\p\Phi}{\p r}}=-\frac{qA}{\epsilon_0}\exp\left\{-\beta q\Phi-\frac{1}{2}\eta B_zr^2-\eta c_{\Psi}\right\}.
\end{equation}
This equation admits the solution 
\begin{equation}
\Phi=-\frac{\eta B_z}{2\beta q}r^2+c_{\Phi}=-\frac{\eta}{\beta q}\Psi+\frac{\eta c_{\Psi}}{\beta q}+c_{\Phi},\label{phistraight}
\end{equation}
such that $\Phi\lr{0}=c_{\Phi}\in\mathbb{R}$ and $\Phi'\lr{0}=0$ with $\Phi'=\p\Phi/\p r$, while the constant $c_{\Psi}$ is determined by the equation
\begin{equation}
\exp\left\{-\eta c_{\Psi}\right\}=\frac{2\eta B_z\epsilon_0}{\beta q^2 A}\exp\left\{\beta q c_{\Phi}\right\}.
\end{equation}
Notice that the corresponding spatial density is constant, $n=A\exp\left\{-\beta q c_{\Phi}-\eta c_{\Psi}\right\}$.
From \eqref{vphistraight} and \eqref{phistraight} it therefore follows that 
the plasma rigidly rotates around the $z$-axis with velocity 
\begin{equation}
v_{\varphi}=-\frac{\eta}{\beta q}r.
\end{equation}
This result also implies that the frequency of the rotation is $\omega_z=-\eta/\beta q$, and that the magnetic field generated by the rotating plasma works to cancel the external magnetic field. 
Furthermore, the conservation of the total magnetic flux, which amounts to the conservation of the ensemble average of the squared radial position of charged particles, $\langle r^2\rangle$, provides radial confinement to a  system with particles initially contained within a given radius.   
It is also worth noticing that this same confinement principle works even for a neutral plasma as long as the third adiabatic invariant holds. 
To see this, consider the simple case $A_p=A_e=A$, $\beta_p=\beta_e=\beta$ and $\eta_p=\eta_e=\eta$.
Then, the Poisson equation for the electrostatic potential admits the trivial solution
$\Phi=0$ such that $n_p=n_e\propto \exp\left\{-\eta\Psi\right\}=\exp\left\{-\eta B_z r^2-\eta c_{\Psi}\right\}$,
which results in radial confinement of the plasma.
Unfortunately, it is known that standard neutral plasmas are poorly confined 
by a straight magnetic field due to the inherent fragility of the third adiabatic invariant, 
which is rapidly destroyed by symmetry breaking electromagnetic perturbations,
and the impossibility of containing the plasma at the vertical ends of the trap via electric fields \cite{Dubin,Malmberg2}.
We remark that, however, we are not aware of positron-electron experiments in this context.


\begin{figure}[h!]
\hspace*{-0cm}\centering
    \includegraphics[scale=0.45]{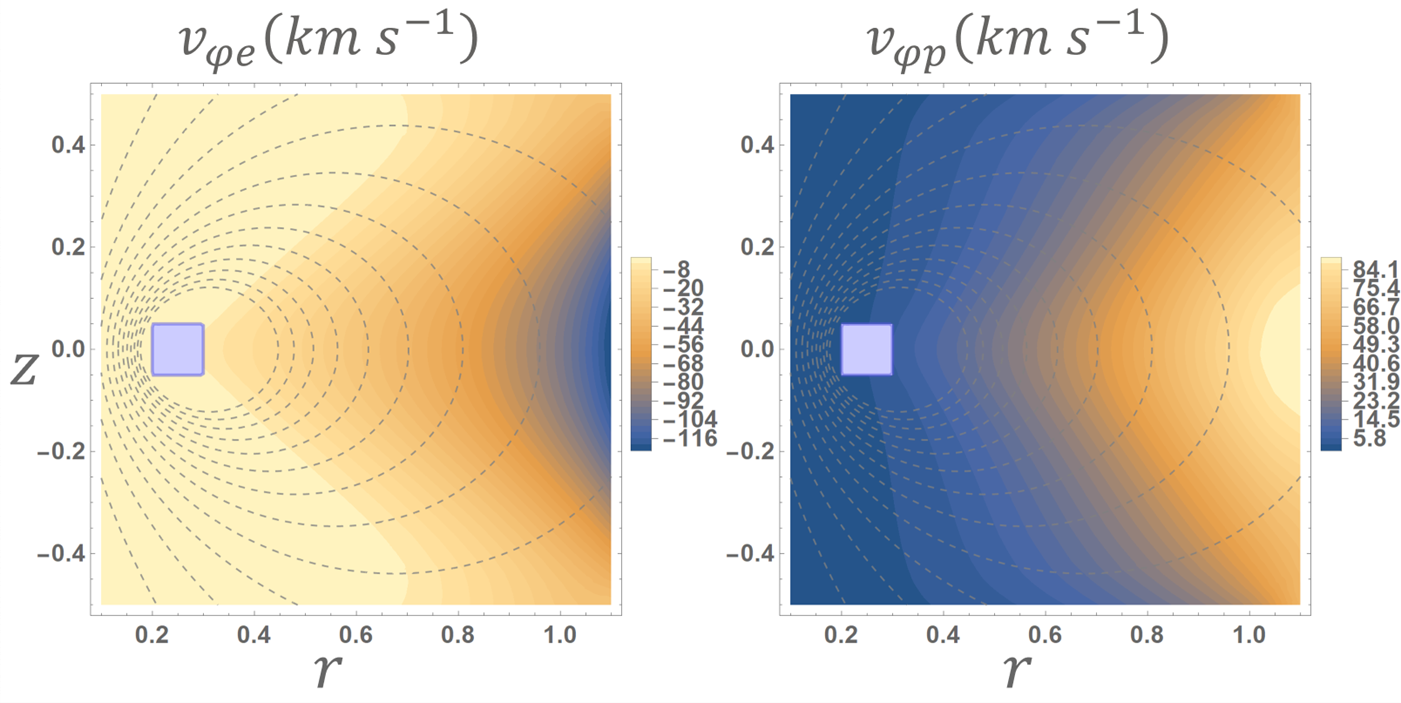}
\caption{\footnotesize Contour plot of the toroidal rotation velocity $v_{\varphi}$ obtained by numerical evaluation of \eqref{vphi} for the
case reported in the second row of figure \eqref{fig4}. Left: the electron toroidal rotation velocity $v_{\varphi e}$ in units of $km\,s^{-1}$. Right: the positron toroidal rotation velocity $v_{\varphi p}$ in units of $km\,s^{-1}$. Dashed contours represent magnetic field lines.}
\label{fig7}
\end{figure}

In a dipole magnetic field 
the situation is essentially different because the confining mechanism is provided by the first adiabatic invariant $\mu$, which forces particles in regions of high magnetic field strength, and not by the magnetic flux $\Psi$ as in a straight magnetic field. As shown in section 2 this also implies that, in contrast with a straight magnetic field, a
dipole magnetic field is  suitable to trap both neutral and nonneutral plasmas. 
Notice also that the toroidal rotation velocity \eqref{vphi} now depends both on the electrostatic potential $\Phi$ and the underlying magnetic field geometry through $\nabla\Psi$, $\nabla\ell$, and $\p_{\ell}^2$.   
Figure \eqref{fig7} shows the toroidal rotation velocity 
obtained by numerical evaluation of \eqref{vphi} for the
case reported in the second row of figure \eqref{fig4}. 
Both the positron toroidal velocity $v_{\varphi p}$
and the electron toroidal velocity $v_{\varphi e}$
increase with $r$, but have opposite directions. 
This implies that, for the case considered, the 
charge dependent drift $v_{\bol{k}}$ is dominant with respect to $\bol{v}_{\bol{E}}$. 
Furthermore, there is a net toroidal current density $J_{\varphi}=e\lr{n_p v_{\varphi p}-n_e v_{\varphi e}}$.
Nevertheless, the resulting magnetic field $B'$ is negligible with respect to the dipole magnetic field. Indeed,
\begin{equation}
B'\approx \mu_0  e R n v_{\varphi}\approx 10^{-9}\, T,
\end{equation}
where $\mu_0$ is the vacuum permeability and the characteristic values $R=1\,m$, $n\approx 10^{11}\, m^{-3}$, and $v_{\varphi}\approx 10^{5}\,m s^{-1}$ were used.

\
\section{Concluding remarks}

In this work, we studied the 
maximum entropy states 
of a collisionless  positron-electron plasma trapped by a dipole magnetic field 
with the aim of elucidating the confinement properties of the system for different ranges of physical parameters. 
Such dipole magnetic field trap has several potential applications, including containment of pair and antimatter plasmas,
experimental investigation of exotic and astrophysical plasmas, as well as technology development such as realization of coherent gamma ray lasers. 

In a dipole magnetic field, the nature of plasma equilibria depends on the presence of adiabatic invariants. 
For such conserved quantities to hold, the time scale of electromagnetic fluctuations affecting the energy of a charged particle
must be longer than the time scale of the periodic motion associated with each adiabatic invariant. 
Each adiabatic invariant constrains the maximum entropy state, resulting in a departure from standard Maxwell-Boltzmann statistics of an ideal gas.

Compared to a plasma trap with a straight magnetic field  
(such as a Penning-Malmberg trap) where radial confinement is provided 
by the conservation of the fragile third adiabatic invariant (the canonical momentum $p_{\varphi}\approx q\Psi$), 
in a dipole magnetic field containment is realized through the first adiabatic invariant (the magnetic moment $\mu$),
which results in a tendency of each charged species to move toward regions of high magnetic field strength $B$.   
For this reason, a dipole magnetic field is suitable to confine both neutral and nonneutral plasmas.
The effect of the second adiabatic invariant (the bounce action $J_{\parallel}$) is to squeeze the
spatial densities of both positrons and electrons along the equatorial line, with the formation of 
characteristic radiation belt like structures around the coil. 

By solving Poisson's equation for the electrostatic potential with the charge density obtained from the maximum entropy states 
as source term, we put the theoretical model to the test, and showed efficient confinement of both 
species for a wide range of physical parameters. The equilibrium profiles appear to be mostly sensible to
asymmetries in the number densities of the two species, and to significant changes in the 
coil potential. This latter fact suggests that the capability to control the coil potential would be a 
desirable property of any experimental dipole magnetic field trap design. 

\section*{Acknowledgment}
The research of NS was partially supported by JSPS KAKENHI Grant No. 21K13851 and 22H04936. 
The author is grateful to H. Saitoh, 
who made helpful suggestions and criticized 
a preliminary draft of the paper.



\end{document}